\documentstyle[12pt]{article}

\input{epsf}

\begin{document}
\baselineskip=.285in

\title{Conformal Phase Transition in Gauge Theories}
\author{
{\sc V. A. Miransky}\\
{\it Institute for Theoretical Physics}\\
{\it 252143 Kiev, Ukraine}\\
{\it and}\\
{\it Department of Physics, Nagoya University}\\
{\it Nagoya 464-01, Japan}\\
\\
{\sc Koichi Yamawaki}\\
{\it Department of Physics, Nagoya University}\\
{\it Nagoya 464-01, Japan}\\
 }
\date{}

\maketitle

\vspace*{-11.5cm}
\begin{flushright}
DPNU-96-58\\
November 1996
\end{flushright}

\vspace*{8.5cm}

\begin{abstract}
The conception of the conformal phase transiton (CPT), which is
relevant for the description of non-perturbative dynamics in
gauge theories, is introduced and elaborated. The main features of
such a phase transition are established. In particular, it is 
shown that in the CPT there is an abrupt change of the spectrum of
light excitations at the critical point, though the phase transition
is continuous. The structure of the effective action describing the
CPT is elaborated and its connection with the dynamics of the partially
conserved dilatation current is pointed out. The applications of these
results to QCD, models of dynamical electroweak symmetry breaking, and
to the description of the phase diagram in (3+1)-dimensional $ SU(N_c)$
gauge theories are considered.  
\end{abstract}

\section{Introduction}

The standard framework for the description of continuous phase transitions 
is the Landau-Ginzburg, or $\sigma$-model like, effective action \cite{1}. 
In particular, in that approach, a phase transition is governed 
by the parameter
\begin{equation}
M^{(2)} \equiv \frac{d^2 V}{d X^2} \left. \right|_{X = 0} \; , \label{1}
\end{equation}
where $V$ is the effective potential and $X$ is an order parameter 
connected with the phase transition. 
When $M^{(2)} > 0 \quad (M^{(2)} <0) $, the symmetric (non-symmetric) phase 
is realized. 
The value $M^{(2)} = 0$ defines the critical point. 

Thus, as $M^{(2)}$ changes, one phase smoothly transforms into another. 
In particular, masses of light excitations are continuous 
(though non-analytic at the critical point) functions of 
such parameters as coupling constants, temperature, etc. 

If $M^{(2)} \equiv 0$, the parameter 
$M^{(4)} \equiv \frac{d^4 V}{d X^4} \left. \right|_{X = 0}$ 
plays the role of $M^{(2)}$, etc.

In this paper, we will describe a non-$\sigma$-model like, 
though continuous, phase transition, which is relevant 
for the description of non-perturbative dynamics in gauge field theories. 
Because, as will become clear below, this phase transition is intimately
connected with a nonperturbative breakdown of the conformal symmetry,
we will call it 
the conformal phase transition (CPT). 

In a $\sigma$-model like phase transition, around the crirical point	
$z=z_c$ (where $z$ is a generic notation for parameters of a theory, 
as the coupling constant $\alpha$, number of particle flavors $N_f$, etc), 
an order parameter $X$ is 
\begin{equation}
X = \Lambda f(z) \label{2}
\end{equation}
($\Lambda$ is an ultraviolet cutoff),
where $f(z)$ has such a non-essential singularity at $z=z_c$
that $\lim f(z)=0$ as $z$ goes to $z_c$ both in symmetric and
non-symmetric phases. The standard form for $f(z)$ is 
$f(z) \sim (z - z_c)^{\nu}$, $\nu > 0$, around $z = z_c$.  
\footnote {
Strictly speaking, Landau and Ginzburg considered the mean-field 
phase transition with $\nu = 1/2$.
By a $\sigma$-model like phase transition, we
understand a more general class, when fields may have anomalous
dimensions \cite{3}. 
}

The CPT is a very different continuous phase transition. We define it
as a phase transition in which an order parameter $X$ is given by
Eq. (\ref{2})
where $f(z)$ has such an $\underline{\mbox{essential}}$ singularity 
at $z=z_c$ that while 
\begin{equation}
\lim_{z \to z_c} f(z) =0 \label{3}
\end{equation}
as $z$ goes to $z_c$ from the side of the non-symmetric phase, 
$\lim f(z) \ne 0 $ 
as $z \to z_c$ from the side of the symmetric phase 
(where $X \equiv 0$). 
Notice that since the relation (\ref{3}) ensures that 
the order parameter $X \to 0$ as $z \to z_c$,
the phase transition is continuous.

There actually exist well-known models in which such a phase
transition is realized.  
As an example of the CPT is the phase transition at $\alpha^{(0)} =0 $ 
($\alpha^{(0)} = \frac{(g^{(0)})^2}{4\pi}$ is the bare coupling constant) 
in massless QCD with a small, say, $N_f \leq 3$, number of fermion flavors. 
In this case, the order parameter $X$, describing chiral symmetry breaking, 
is $X \sim \Lambda_{\mbox{QCD}}$ and 
\begin{equation}
X \sim \Lambda_{\mbox{QCD}} \sim \Lambda f(\alpha^{(0)}) \; , \label{4}
\end{equation}
where $ f(\alpha^{(0)}) \simeq \exp \left( -\frac{1}{b \alpha^{(0)}} \right)$ 
($b$ is the first coefficient of the QCD $\beta$ function). 
The function $ f(\alpha^{(0)}) $ goes to zero only if $\alpha^{(0)} \to 0$ 
from the side of $Re \alpha^{(0)} > 0$. 

The above example is somewhat degenerate: 
the critical point $\alpha^{(0)}_c =0$ is at the edge of the physical space 
with $ \alpha^{(0)} \geq 0$. 
A more regular example of the CPT is given by the phase transition at 
$g^{(0)} = 0 $ in the $(1+1)$-dimensional Gross-Neveu model: 
in that case both positive and negative values of $g^{(0)}$ are physical 
(see Sec.3). 

There may exist more sophisticated realizations of the CPT. 
As will be discussed in Sec.7, an example of the CPT may be provided 
by the phase transition with respect to the number of fermion flavors $N_f$ 
in a $SU(N_c)$ vector-like gauge theory in $(3+1)$ dimensions, considered 
by Banks and Zaks long ago \cite{2}. 
In that case, unlike the phase transition at $\alpha^{(0)}=0$ in QCD, 
the critical value $N_f^{cr}$ separates two physical phases, 
with $N_f < N_f^{cr}$ and $N_f \geq N_f^{cr}$. 

There may exist other examples of the CPT. 
Also there may exist phase transitions in $(2+1)$-dimensional theories 
which ``imitate'' the CPT (see Sec.5).

The main goal of this paper is to reveal the main features of the
CPT (common for its different realizations) and to apply the conception 
of the CPT to concrete models.  

The CPT is $\underline{\mbox{not}}$ 
a $\sigma$-model like phase transition, though it is continuous. 
In particular, in the CPT, one cannot introduce the parameters 
$M^{(2n)} = \frac{d^{2n} V}{d X^{2n}} \left. \right|_{X=0} \; , 
n=1,2, \cdot \cdot \cdot \; , $ governing the phase transition. 
Another characteristic feature of the CPT is an abrupt
change of the number of 
light excitations as the critical point is crossed 
(though the phase transition is continuous). While evident in QCD
and the Gross-Neveu model, it is realized in a more subtle way
in the general case. 
This feature implies a specific form of the effective
action describing
light excitations in theories with the CPT, which will be discussed
in this paper.

The paper is organized as follows. In Sec.2 we consider the 
properties of the spectrum of light excitations around the critical
point $z=z_c$ in the CPT. We show that, though the CPT is a continuous
phase transition, there is an abrupt change of the spectrum of light
excitations as the critical point is crossed. In Sec.3 we describe
the chiral phase transition in the $D$-dimensional
($2 \leq D<4$) Nambu-Jona-Lasinio (Gross-Neveu) model. We study the
CPT in the 2-dimensional Gross-Neveu model. This allows to illustrate
main features of the CPT in a very clear way. In Sec.4 we study the
CPT in quenched QED4, which is relevant for the study of the phase
transition with respect to fermion flavors in a 4-dimensional
$SU(N_c)$ gauge theory. In Sec.5 the main features of the CPT
are summarized and the realization of a pseudo-CPT in QED3 is
considered. In Sec.6 the structure of the effective action in
theories with the CPT and the realization of the dynamics of
the partially conserved dilatation current in these theories
are discussed.
In Sec.7 the phase diagram with respect to the bare
coupling constant $ {\alpha}^{(0)} $ and the number of fermion
flavors $N_f$ in a $SU(N_c)$ gauge theory is considered. In
particular, we suggest a modified (as compared to that suggested in
Ref. [3]) phase diagram. Possibilities of the examination of this
phase diagram in lattice computer simulations are discussed. In Sec.8
we summarize the main results of the paper.
In the 
Appendix some useful relations are derived.

\section{Peculiarities of the Spectrum of Light Excitations in the CPT}

As was already pointed out in Introduction, in the
case of the $\sigma$-model 
like phase transition, masses of light excitations are continuous functions 
of the parameters $z$ around the critical point $z=z_c$ 
(though they are non-analytic at $z=z_c$). 
Let us show that the situation in the case of the CPT is different: 
there is an abrupt change of the spectrum of light excitations, 
as the critical point $z=z_c$ is crossed. 

Let us start from a particular, and important, case of the CPT connected 
with dynamical chiral symmetry breaking. In this case, in the non-symmetric 
phase, amongst light (with masses much less than cutoff $\Lambda$) 
excitations, there are massless Nambu-Goldstone (NG) bosons $\pi$, 
their chiral partners, $\sigma$ bosons, and light (with $m_{dyn} \ll \Lambda$) 
fermions. The masses of $\sigma$ and fermions are given by scaling relations:
\begin{eqnarray}
M_{\sigma}^2 &=& C_{\sigma} \Lambda^2 f(z) \label{5} \\
m_{dyn}^2 &=& C_{f} \Lambda^2 f(z) \; , \label{6} 
\end{eqnarray}
where $C_{\sigma}$ and $C_f$ are some positive constants, and $f(z)$ is a 
universal scaling function. 
Because of the assumption (\ref{3}), $M_{\sigma}^2$ and
$m_{dyn}^2$ are 
indeed much less than $\Lambda^2$, when $z$ is near $z_c$ from the 
side of the non-symmetric phase. 

Now, are there light $\pi$ and $\sigma$ resonances in the symmetric phase, 
with $m_{dyn}=0$? 
Since, as was assumed, $\lim f(z) \ne 0 $ as $z \to z_c$ in that phase, one 
should expect that there are no light resonances. 
Let us show that this is indeed the case.

One might think that in the symmetric phase the mass relation for 
$\pi$ and $\sigma$ is yielded by the analytic continuation of the relation 
(\ref{5}) for $M_{\sigma}^2$.
However this is not the case. The point is that while in 
the non-symmetric phase, $\pi$ and $\sigma$ bosons are described by 
Bethe-Salpeter (BS) equations with a non-zero fermion mass, in the symmetric 
phase they are described by BS equations with $m_{dyn} \equiv 0$. 
Because of that, BS equations (and, more generally, all the Schwinger-Dyson 
equations for Green's functions) in the symmetric phase are not yielded 
by an analytic continuation of the equations in the non-symmetric phase. 

To overcome this obstacles, we shall use the following trick. 
In the non-symmetric phase, besides the stable solution with 
$m_{dyn} \ne 0$, there is also an unstable solution with $m_{dyn} = 0$. 
In that solution, $\pi$ and $\sigma$ bosons are tachyons: 
$M_{\pi}^2=M_{\sigma}^2 \equiv M_{tch}^2<0$. 
Since the replacement of $m_{dyn} \ne 0 $ by $m_{dyn}=0$ 
(at fixed values of the parameters $z$) does not change the ultraviolet 
properties of the theory, the scaling relation for the tachyon masses has 
the same form as that in Eqs. (\ref{5}) and (\ref{6}): 
\begin{equation}
M_{\pi}^2=M_{\sigma}^2 = M_{tch}^2 =
-C_{tch} \Lambda^2 f(z), \quad C_{tch}>0. 
\label{7} 
\end{equation}
Since now $m_{dyn} = 0$, the BS equations for tachyons have the same form 
as the BS equations for $\pi$ and $\sigma$ in the symmetric phase; 
the difference between these equations is only in the values of $z$ 
(for convenience, we shall assume that $z > z_c \quad (z < z_c)$ in 
the non-symmetric (symmetric) phase). 
Then, in the symmetric phase,
\begin{equation}
M_{\pi}^2=M_{\sigma}^2 = -C_{tch} \Lambda^2 f(z) \; ; C_{tch}>0, 
\label{8} 
\end{equation}
with $z<z_c$ and $C_{tch}$ from Eq. (\ref{7}). 
Notice that because in the symmetric phase $\pi$ and $\sigma$ bosons 
decay to massless fermions and antifermions, $M_{\pi}^2$ and $M_{\sigma}^2$ 
are complex, i.e. $\pi$ and $\sigma$ are now resonances, 
if they exist at all. 

Since, by definition, in the CPT, $\lim f(z) \ne 0 $ as $z \to z_c -0 $, 
we conclude from Eq. (\ref{8}) that there are no light resonances near 
the critical point from the side of the symmetric phase: 
$
| M_{\pi}^2 |= |M_{\sigma}^2| \sim \Lambda^2 
$ as $z \to z_c -0 $. 

So far, for concreteness, we have considered the case of 
dynamical chiral symmetry breaking. But it is clear that 
(with minor modifications) this consideration can be extended to 
the general case of the CPT connected with spontaneous breakdown of 
other symmetries. 

Notice also that the relation in Eq. (\ref{8}) can be useful
for general phase 
transitions and not just for the CPT. 
The point is that the scaling function $f(z)$ can be determined from 
the gap equation for the order parameter ($m_{dyn}^2$, in the case of 
chiral symmetry) which is usually much simpler than the BS equation for 
massive composites. For example, 
an abrupt change of the spectrum at the critical point $z=z_c$ have been 
revealed in some models: in quenched QED4 \cite{4,5} and QED3 \cite{6}.
This conclusion was based on an analysis of the effective action
\cite{4} and the BS equation \cite{5,6}, considered in a rather crude
approximation. On the other hand, since the determination of the
scaling function $f(z)$ in these models is a much simpler task, this
conclusion can be firmly established in the present approach (see
Secs.4 and 5). 
Thus the present consideration yields a simple and general criterion of 
such a peculiar behavior of the spectrum of light excitations. 

It is clear that the abrupt change of the spectrum discussed above 
implies rather peculiar properties of the effective action for 
light excitations at the critical point. Below we shall consider 
this problem in more detail. We shall also reveal 
an intimate connection between this point and the essential difference 
of the character of the breakdown of the conformal symmetry 
in different phases of theories with the CPT.

\section{$D$-dimensional Nambu-Jona-Lasinio (Gross-Neveu) model. 
The CPT at D=2}

In this section we consider the dynamics in the $D$-dimensional 
$(2 \leq D < 4)$ Nambu-Jona-Lasinio (Gross-Neveu) model and, in particular, 
describe the CPT in the Gross-Neveu (GN) model at $D=2$. 
This will allow to illustrate main features of the CPT in a very clear way. 

The Lagrangian density of the $D$-dimensional GN model, with the 
$U(1)_L \times U(1)_R$ chiral symmetry, is 
\begin{equation}
{\cal L}= \frac{1}{2} \left[ \bar{\psi}, (i \gamma^{\mu}\partial_{\mu}) \psi 
\right] 
+ \frac{G}{2} \left[ (\bar{\psi} \psi)^2 + (\bar{\psi} i \gamma_5 \psi)^2 
\right] , \label{9}
\end{equation}
where $\mu=0,1, \cdot \cdot ,D-1$, and the fermion field carries 
an additional ``color'' index $\alpha = 1,2, \cdot \cdot , N_c$. 
The theory is equivalent to the theory with the Lagrangian density 
\begin{equation}
{\cal L}' = \frac{1}{2} \left[ \bar{\psi}, (i\gamma^{\mu}\partial_{\mu}) 
\psi \right] -\bar{\psi} (\sigma + i \gamma_5 \pi) \psi 
- \frac{1}{2G} (\sigma^2 + \pi^2) . \label{10}
\end{equation}
The Euler-Lagrange equations for the auxiliary fields $\sigma $ and $\pi$ 
take the form of constraints:
\begin{equation}
\sigma = -G \bar{\psi}\psi \; , \pi = - G \bar{\psi}i\gamma_5 \psi, \label{11}
\end{equation}
and the Lagrangian density (\ref{10}) reproduces Eq. (\ref{9}) upon 
application of the constraints  (\ref{11}). 
The effective action for the composite fields $\sigma $ and $\pi$ is 
obtained by integrating over fermions in the path integral:
\begin{equation}
\Gamma(\sigma , \pi) = -i \mbox{Tr Ln } [i\gamma^{\mu}\partial_{\mu} - 
(\sigma + i \gamma_5 \pi) ]  
- \frac{1}{2G} \int d^D x (\sigma^2 + \pi^2). \label{12}
\end{equation}
The low energy dynamics are described by the path integral 
(with the integrand $\exp(i\Gamma) $ ) over the fields $\sigma$ and $\pi$. 
As $N_c \to \infty$, the path integral is dominated by the stationary points 
of the action: $\frac{\delta \Gamma}{\delta \sigma} = 
\frac{\delta \Gamma}{\delta \pi} = 0.$ 

Let us look at the effective potential in this theory. It is \cite{7}
\begin{equation}
V(\sigma , \pi) = \frac{4 N_c \Lambda^D}{(4\pi)^{D/2} \Gamma(D/2)}
\left[ ( \frac{1}{g} - \frac{1}{g_{cr}}) \frac{\rho^2}{2\Lambda^2} 
+ \frac{2}{4-D} \frac{\xi_D}{D} (\frac{\rho}{\Lambda})^D \right] 
+ O(\frac{\rho^4}{\Lambda^4}) , \label{13}
\end{equation}
where $\rho = (\sigma^2 + \pi^2 )^{1/2}, \xi_D = B(D/2-1,3-D/2),$ 
the dimensionless coupling constant $g$ is 
\begin{equation}
g = \frac{4 N_c \Lambda^{D-2}}{(4\pi)^{D/2}\Gamma(D/2)} G, \label{14}
\end{equation}
and the critical coupling $g_{cr} = \frac{D}{2}-1$. 

At $D>2$, one finds that 
\begin{equation}
M^{(2)} \equiv \frac{d^2 V}{d \rho^2} \left. \right|_{\rho=0} \simeq 
\frac{4 N_c \Lambda^{D-2}}{(4\pi)^{D/2}\Gamma(D/2)} \frac{g_{cr}-g}{g_{cr}g}. 
\label{15}
\end{equation}
The sign of $M^{(2)}$ defines two different phases: 
$M^{(2)} > 0$ $(g < g_{cr})$ corresponds to the symmetric phase and 
$M^{(2)} < 0$ $(g > g_{cr})$ corresponds to the phase with spontaneous chiral 
symmetry breaking, $U(1)_L \times U(1)_R \to U(1)_{L+R}$. 
The value $M^{(2)} =0$ defines the critical point $g=g_{cr}$. 

Therefore at $D>2$, a $\sigma $-model like phase transition is realized.
However the case $D=2$ is special: 
now $g_{cr} \to 0$ and $\xi_D \to \infty$ as $D \to 2.$ In this case 
the effective potential is the well-known potential
of the Gross-Neveu model \cite{8}: 
\begin{equation}
V(\sigma , \pi) = \frac{N_c}{2\pi g} \rho^2 - \frac{N_c \rho^2 }{2\pi}
 \left[ \ln \frac{\Lambda^2}{\rho^2} + 1 \right]. \label{16}
\end{equation}
The parameter $M^{(2)}$ is now : 
\begin{equation}
M^{(2)} = \frac{d^2 V}{d \rho^2} \left. \right|_{\rho=0} \to + \infty. 
\end{equation} 
Therefore, in this model, one cannot use $M^{(2)}$ as a parameter governing 
the continuous phase transition at $g = g_{cr} =0$ : 
the phase transition is not a $\sigma$-model like phase transition 
in this case. 
Indeed, as follows from Eq. (\ref{16}), 
the order parameter, which is a solution to the gap equation 
$\frac{d V }{d \rho} =0$, is 
\begin{equation}
\bar{\rho} = \Lambda \exp ( -\frac{1}{2g}) \label{18}
\end{equation}
in this model. The function $f(z)$, defined in Eq. (\ref{2}), is now 
$f(g) = \exp (- \frac{1}{2g})$, i.e., $z=g$, and therefore the CPT 
takes place in this model at $g=0$: $f(g)$ goes to zero only if
$g \to 0$ from the side of the non-symmetric phase. 

Let us discuss this point in more detail. 

At $D \geq 2$, the spectrum of the $\sigma$ and $\pi$ excitations 
in the symmetric solution, with $\bar{\rho}=0$, is defined by the following 
equation (in leading order in $\frac{1}{N_c}$) \cite{7}: 
\begin{equation}
(\frac{1}{g} - \frac{1}{g_{cr}}) \Lambda^{D-2} + \frac{\xi_D}{2-D/2} 
(-M_{\pi}^2)^{D/2-1} = 0. \label{19}
\end{equation}
Therefore at $D>2$, there are tachyons with 
\begin{equation}
M_{\pi}^2=M_{\sigma}^2 = M^2_{tch} = 
-\Lambda^2  (\frac{4 -D}{2\xi_D})^{\frac{2}{D-2}} 
(\frac{g-g_{cr}}{g_{cr}g})^{\frac{2}{D-2}} \label{20} 
\end{equation}
at $g > g_{cr}$, and  at $g < g_{cr}$ there are ``resonances'' with 
\begin{equation}
|M_{\pi}^2| = |M_{\sigma}^2| = 
\Lambda^2 (\frac{4 -D}{2\xi_D})^{\frac{2}{D-2}} 
(\frac{g_{cr}-g}{g_{cr}g})^{\frac{2}{D-2}} \; , \label{21} 
\end{equation} 
which agrees with Eq. (\ref{8}).
\footnote{
For our purposes, it is sufficient to calculate the absolute value of 
$M_{\pi}^2$. Notice that, as follows from Eq. (\ref{19}), narrow resonances 
occur near $D=4$: $\frac{\Gamma}{M_R} \simeq \pi \frac{4-D}{D-2} \quad 
(M_{\pi} = M_R - i \frac{\Gamma}{2} )$ . 
} 
Eq. (\ref{21}) implies that the limit $D \to 2$ is special. 
One finds from Eq. (\ref{19}) that at $D=2$
\begin{equation}
M_{\pi}^2=M_{\sigma}^2 = M^2_{tch} = - \Lambda^2 \exp ( - \frac{1}{g}) 
\label{22}
\end{equation}
at $g> 0$, and 
\begin{equation}
|M_{\pi}^2| = |M_{\sigma}^2| =  \Lambda^2 \exp (  \frac{1}{|g|}) 
\label{23}
\end{equation}
at $g< 0$, i. e., in agreement with the main feature of the CPT, 
there are no light resonances in the symmetric phase at $D=2$. 

The effective potential (\ref{16}) can be rewritten as 
\begin{equation}
V(\sigma , \pi) = \frac{N_c \rho^2}{2\pi}
 \left[ \ln \frac{\rho^2}{\bar{\rho}^2}- 1 \right]  \label{24}
\end{equation}
(with $\bar{\rho}$ given by Eq. (\ref{18})) in the non-symmetric phase. 
That is, in this phase $V(\sigma , \pi)$ is finite in the continuum limit 
$\Lambda \to \infty$ after the renormalization of the coupling constant, 
\begin{equation}
g=\frac{1}{\ln \frac{\Lambda^2}{\bar{\rho}^2}} \label{25}
\end{equation}
(see Eq. (\ref{18})). But what is the form of the effective potential 
in the continuum limit in the symmetric phase, with $g<0$ ? 
As Eq. (\ref{16}) implies, it is infinite as $\Lambda \to \infty$ : 
indeed at $g<0$, there is no way to cancel the logarithmic divergence in $V$. 

It is unlike the case with $D> 2$ : in that case, using Eq. (\ref{15}), 
the potential (\ref{13}) can be put in a $\sigma$-model like form : 
\begin{equation}
V(\sigma , \pi) = \frac{M^{(2)}}{2} \rho^2 
+ \frac{8 N_c}{(4\pi)^{D/2} \Gamma(D/2)} \frac{\xi_D}{(4-D)D} \rho^D. 
\label{26}
\end{equation}
However, since $M^{(2)} = \infty$ at $D=2$, the $\sigma$-model like form 
for the potential is not available in the Gross-Neveu model. 

What are physical reasons of such a peculiar behavior of the effective 
potential at $D=2$ ? 
Unlike the case with $D>2$, at $D=2$ the Lagrangian density 
(\ref{9}) defines a conformal theory in the classical limit. 
By using the conventional approach, one can derive the following 
equation for the conformal anomaly in this model (see the Appendix): 
\begin{equation}
\partial^{\mu} D_{\mu} = \theta_{\mu}^{\mu} = 
\frac{\pi}{2 N_c} \beta (g)  \left[ (\bar{\psi}\psi)^2 + (\bar{\psi} i 
\gamma_5 \psi )^2 \right] \label{27} 
\end{equation}
where $D_{\mu}$ is the dilatation current, $\theta_{\nu}^{\mu}$ 
is the energy-momentum tensor, and the $\beta$ function 
$
\beta = \frac{\partial g}{\partial \ln \Lambda}
$. 
It is $\beta (g) = -g^2 $ both in the non-symmetric and 
symmetric phases. 
While the non-symmetric phase corresponds to asymptotically free 
dynamics, the symmetric phase (with $g < 0$)
defines infrared free dynamics : 
as $\Lambda \to \infty$, we are led to a free theory of massless 
fermions, which is of course conformal invariant. 

On the other hand, in the non-symmetric phase the conformal symmetry 
is broken, even as $\Lambda \to \infty$. 
In particular, Eq. (\ref{24}) implies that
\begin{equation}
\langle 0 | \theta_{\mu}^{\mu} | 0 \rangle = 4V(\bar{\rho}) =
- \frac{2N_c}{\pi}{\bar{\rho}}^2 \ne 0
\label{28}
\end{equation}
in leading order in $\frac{1}{N_c}$ in that phase. 

The physics underlying this difference between the two phases is clear : 
while negative $g$ correspond to repulsive interactions between 
fermions, attractive interactions at positive $g$ lead to the formation of 
bound states, thus breaking the conformal symmetry. 

Notice the following interesting point. As follows from Eq. (\ref{26}), 
at $D > 2$ the conformal symmetry is broken by a relevant 
(superrenormalized) mass operator: 
its dynamical dimension is $d=2$ at all $ 2 \leq D \leq 4$. 
On the other hand, at $D=2$ the symmetry is broken by a marginal 
(renormalized) operator with the dynamical dimension $d=2$. 
This point is reflected in that while at $D=2$ the expression for 
the order parameter $\bar{\rho}$ has an essential singularity at the 
critical point $g=g_{cr}=0$, at $D>2$, the singularity at 
$g=g_{cr}$ in $\bar{\rho}$ is not essential : 
as follows from Eq. (\ref{13}), the solution to the gap equation 
$\frac{d V}{d \rho} = 0 $ is $ \bar{\rho} \sim \Lambda 
(g-g_{cr})^{\frac{1}{D-2}}$ 
in that case. As is known, the essential singularity implies the absence of 
of fine tuning for bare parameters. This is another reason 
why the CPT is so interesting. 

Thus the CPT, in accordance with its name, describes the two essentially
different realizations of the conformal symmetry in the symmetric and
non-symmetric phases.

If one adds a fermion mass term, $m^{(0)} \bar{\psi} \psi $, 
in the 2-dimensional GN model, the conformal and chiral symmetries will 
be of course broken in both phases. 
However, there remains an essential trace of the CPT also in this case : 
an abrupt change of the spectrum of light excitations still takes place. 
While now in the subcritical $(g< g_{cr}=0)$ phase repulsive interactions
between massive fermions are 
take place (and there are no light resonances there), in the supercritical 
$(g > g_{cr} =0)$ phase the PCAC dynamics, describing interactions between 
fermions and light $\pi$ and $\sigma$ bosons, is realized. 
\footnote{
We are of course aware that the exact solution in the non-symmetric phase 
of the 2-dimensional GN model yields a realization of 
the Berezinsky-Kosterlitz-Thouless(BKT) phase : 
though chiral symmetry is unbroken, the parameter $\bar{\rho}$ still 
defines the fermion dynamical mass, and the would-be NG boson $\pi$ 
transforms into a BKT gapless excitation \cite{9}. 
} 

Besides the point that in the 2-dimensional GN model both 
subcritical and supercritical phases are physical, this picture
is similar to that in QCD. It is hardly surprising: in both models 
the dynamics in the supercritical phases are
asymptotically free. We will however
argue that the main features
of the CPT found in the GN model will retain valid (with 
appropriate minor modifications) in the general case.

\section{The CPT in quenched QED4}

In this section we shall describe the main features of the CPT
in quenched QED4. The dynamics in this model is relevant for some 
scenarios of dynamical electroweak symmetry breaking and has been
intensively discussed in the literature (for a review see Ref. [10]).
In the present paper the emphasis of the discussion will be on the
points relevant for the general CPT in gauge theories. 

We shall consider the ladder (rainbow) approximation in massless
QED4.  Since the contribution of fermion loops is omitted, the
perturbative $\beta$-function equals zero in this approximation. 
However, as is well known \cite{10,11,12}, beyond the critical value
$\alpha = \alpha_c \sim 1$, there are nonperturbative divergences
which break the conformal symmetry in the model.  Moreover, since at
$\alpha = \alpha_c$, the anomalous dimension $\gamma_m$ of the chiral
operators $\bar{\psi} \psi$ and $\bar{\psi}  i \gamma_5 \psi$ is
$\gamma_m = 1$ \cite{12,13}, the four-fermion operators $(\bar{\psi} \psi)^2$
and $\bar\psi i \gamma_5 \psi)^2$ become (marginally) relevant: 
their dynamical dimension $d$ is $d=
d_{c} - 2 \gamma_m = 4$, where
$d_{c} = 6$ is their canonical dimension.

Therefore, it is appropriate to include these four-fermion
operators in the QED action.  This leads to the gauged
Nambu-Jona-Lasinio model \cite{13}:
\begin{equation}
{\cal L} = - \frac{1}{4} \left(F_{\mu \nu} \right)^2 + \frac{1}{2}
\left[ \bar\psi, (i
\gamma^\mu D_\mu)\psi \right]
+ \frac{G}{2} \left[ (\bar\psi \psi)^2 + (\bar\psi i \gamma_5 \psi)^2
\right], \label{29}
\end{equation}
\noindent where $D_\mu = \partial_\mu - ie A_\mu$ (for simplicity, we consider
the chiral symmetry $U_L(1) \times U_R(1)$).  In this model, the gauge
interactions are treated in the ladder approximation and the
four-fermion interactions are treated in the Hartree-Fock (mean
field) approximation.

Since the coupling constant $G$ is dimensional, one may think that
the four-fermion interactions in Eq. (\ref{29}) explicitly break the
conformal symmetry.  The real situation is however more subtle.  In
Fig. 1, we show the critical line in this model \cite{14}, 
dividing the symmetric
phase, with the unbroken $U_L(1) \times U_R(1)$, and the phase
with the spontaneously broken chiral symmetry $(U_L(1) \times U_R(1)
\rightarrow U_{L+R}(1)$).  Each point of the
critical line corresponds to a continuous phase transition.  We
distinguish two parts of the critical line:
\begin{equation}
g \equiv \frac{G\Lambda^2}{4 \pi^2} = \frac{1}{4} \left[ 1 + \left( 1 -
\frac{\alpha}{\alpha_c} \right)^{1/2} \right]^2, \: \: 
\alpha_c = \frac{\pi}{3}, \label{30}
\end{equation}
\noindent at $g > \frac{1}{4}$, and
\begin{equation}
\alpha = \alpha_c \label{31}
\end{equation}
\noindent at $g < \frac{1}{4}$.  The anomalous dimension $\gamma_m$ of the
operators $\bar\psi \psi$ and $\bar\psi i \gamma_5 \psi$ along the
critical line is \cite{15}
\begin{equation}
\gamma_m = 1 + \left( 1 - \frac{\alpha}{\alpha_c} \right)^{1/2}. \label{32}
\end{equation}
\noindent In this approximation, the anomalous dimension of the
four-fermion operator $\left[ (\bar\psi \psi)^2 + (\bar\psi i
\gamma_5 \psi)^2 \right]$ equals $2 \gamma_m$.  Therefore
while this operator
indeed breaks the conformal symmetry along
the part (\ref{30}) of the critical line,
it is a marginal (scale
invariant) operator along the part of the critical line with $\alpha
= \alpha_c$:  its dynamical dimension is $d_{\bar\psi \psi} = 6 -
2\gamma_m = 4$ there.

Thus the part (\ref{31}) of the critical line with $\alpha = \alpha_c$ is
special.  In this case the symmetric phase is not only chiral
invariant but also conformal invariant.  On the other hand, in the
non-symmetric phase, both these symmetries are broken:  while the
chiral symmetry is broken spontaneously, the conformal symmetry is
broken explicitly (see below) \cite{13,14}.

Unlike the case of the NJL model, where the Lagrangian density 
(\ref{10}) with the auxiliary fields $\sigma$ and $\pi$ was 
used to derive the effective action $\Gamma[\sigma, \pi]$, 
now we will derive another effective action: 
generating functional for proper vertices of the local composite 
operators $\bar{\psi}\psi$ and $\bar{\psi}i \gamma_5 \psi$. 
The point is that the trick, used in Eq. (\ref{10}), to introduce 
the fields $\sigma$ and $\pi$ 
does not work in pure QED ($G=0$). 
It is also unclear how well this effective action describes 
the collective excitations $\sigma$ and $\pi$ in the case of 
weakly coupling four-fermion interactions (when QED forces 
dominate). 
We shall discuss the connection between these two actions below 
(the effective action based on using the auxiliary fields 
$\sigma$ and $\pi$ was considered in this model 
in Refs. \cite{16,17,7}). 

Let us describe the effective action (generating functional for
proper vertices) for the local composite operators $\bar\psi \psi$
and $\bar\psi i \gamma_5 \psi$ in the gauged NJL model.  This
effective action was derived in Ref. [4].  In the present paper, we
will describe in more detail those features of the action which are
relevant for understanding the nature of the CPT.

The effective action is constructed in the standard way.
First, one introduces
a generating functional for Green's functions of the operators
$\hat{\rho}_i (x), \: \: i = 1, 2 \; \; (\hat{\rho}_1 = \bar\psi \psi, \:
\hat{\rho}_2 = \bar\psi i \gamma_5 \psi)$:
$$
Z \left( \{ J_i \} \right)= \exp \left[ i W \left( \{ J_i
\} \right) \right]
$$
\begin{equation}
= \int d \varphi \exp \left[ i \int d^4 x \{ {\cal
L} (x) + \sum_{i=1}^2 J_i (x) \hat{\rho}_i (x) \} \right], \label{33}
\end{equation}
\noindent where the $\varphi$ integration is functional, $J_i(x)$ is 
the source
for $\hat{\rho}_i (x)$, and ${\cal L}(x)$ is the Lagrangian density
(\ref{29}) (the symbol $\varphi(x)$ represents all fields of the model).

The effective action for the operators $\hat{\rho}_i (x)$ is a
Legendre transform of the functional $W \left( \{J_i \} \right)$:
\begin{equation}
\tilde{\Gamma}\left( \{ \rho_i \} \right)
= W \left( \{ J_i \} \right) - \int
d^4 x \sum_{i=1}^2 J_i (x) \rho_i (x), \label{34}
\end{equation}
\noindent where $\rho_i (x) \equiv < 0 \vert \hat{\rho}_i (x) \vert 0
>$ (we use ``tilde'' here in order to distinguish this effective action 
from that for the auxiliary fields $\sigma$ and $\pi$ ). 
>From Eqs. (\ref{33}) and (\ref{34}) one finds that the following 
relations are satisfied:
\begin{eqnarray}
\frac{\delta W}{\delta J_i(x)} &=& \rho_i (x), \label{35} \\
\frac{\delta \tilde{\Gamma}}{\delta \rho_i (x)} &=& - J_i (x). \label{36}
\end{eqnarray}

The effective action $\tilde{\Gamma} \left( \{ \rho_i \} \right)$ can be
expanded in powers of derivatives of the fields $\rho_i (x)$:
\begin{equation}
\tilde{\Gamma} \left( \{ \rho_i \} \right) = \int d^4 x 
\left[ -\tilde{V} \left( \{
\rho_i \} \right) + \frac{1}{2} \tilde{Z}_{ij} \left( \{ \rho_i \} \right)
\partial_\mu \rho_i \partial^\mu \rho_j + ... \right] \label{37}
\end{equation}
\noindent where $\tilde{V} \left( \{ \rho_i \} \right)$ is the effective
potential.

The calculation of
the effective potential is reduced to finding the Legendre transform
of the functional $W \left( \{ J_i \} \right)$ with the sources
$J_i$ independent of coordinates $x$.  Because of chiral symmetry, W
depends on the chiral invariant $\tilde{J}^2 = J_1^2 + J_2^2$. 
Therefore to determine the form of W, it is sufficient to consider
the source term with $J_1 \neq 0, \: J_2 = 0$.  Then, owing to the
relation (\ref{35}), one finds that
\begin{equation}
w(J_1) = \int^{\Sigma_0(J_1)} < 0 \vert \bar \psi \psi \vert 0 >_J
\frac{dJ}{d\Sigma_0} d\Sigma_0, \label{38}
\end{equation}
where
\begin{equation}
W(J_1) = w(J_1) \int d^4 x \label{39}
\end{equation}
and $\left. \Sigma_0 \equiv \Sigma(p^2) \right|_{p^2=0}$.  Here $\Sigma(p^2)$ 
is the
fermion mass function (the fermion propagator is $G(p) = \left(A(p^2)
p^\mu \gamma_\mu - \Sigma(p^2) \right)^{-1}$).

Since the functional $W(J_1)$ corresponds to the source term with
$J_1 \neq 0, J_2 = 0$, the condensate $< 0 \vert \bar\psi \psi \vert
0 >_J$ is related to the gauged NJL model with the bare mass
$m^{(0)} = -J$.

Now we need to know the following information concerning the fermion
propagator in rainbow QED4 (for a review see Ref. \cite{10}).  In the
Landau gauge, the functions $A(p^2)$ and $\Sigma(p^2)$ satisfies the
equations
\begin{equation}
A(p^2) = 1, \label{40}
\end{equation}
$$
\Sigma(p^2) = m^{(0)} + \frac{g}{\Lambda^2} \int _0^{\Lambda^2} dq^2
\frac{q^2 \Sigma(q^2)}{q^2 + \Sigma^2(q^2)} +
$$
\begin{equation}
+ \frac{3 \alpha}{4 \pi} \int_0^{\Lambda^2} dq^2 \frac{q^2}{q^2 +
\Sigma^2(q^2)} \left[ \frac{\theta(p^2-q^2)}{p^2} +
\frac{\theta(q^2-p^2)}{q^2} \right] \Sigma(q^2), \label{41}
\end{equation}
where $g\equiv \frac{G\Lambda^2}{4 \pi^2}$, $\Lambda$ is an
ultraviolet cutoff and $m^{(0)} \equiv -J$ is the bare mass of
fermions (equation (\ref{41}) is written in Euclidean space).

Differentiating Eq. (\ref{41}) with respect to $p^2$, one finds that
$\Sigma(p^2)$ satisfies the differential equation
\begin{equation}
\frac{d}{dp^2} \left[ p^2 \frac{d \Sigma(p^2)}{d p^2} \right] +
\frac{3\alpha}{4 \pi} \frac{\Sigma(p^2)}{p^2 + \Sigma^2(p^2)} = 0 
\label{42}
\end{equation}
and two boundary conditions:
\begin{equation}
\left. m^{(0)} \equiv -J = \left[ \left( 1 + \frac{4 \pi g}{3 \alpha}
\right) p^2 \frac{d\Sigma(p^2)}{dp^2} + \Sigma(p^2) \right] \right|
_{p^2=\Lambda^2}, \label{43}
\end{equation}
\begin{equation}
\lim_{p^2 \rightarrow 0} \left[ (p^2)^2 \frac{d\Sigma(p^2)}{dp^2}
\right] = 0 . \label{44}
\end{equation}
Differentiating Eq. (\ref{41}) with respect to $p^2$ at $p^2 = \Lambda^2$,
we find that the chiral condensate $< 0 \vert \bar\psi \psi \vert )
>_J$ is
$$
< 0 \vert \bar\psi \psi \vert 0 >_J = - \frac{1}{4\pi^2}
\int_0^{\Lambda^2} dq^2 \frac{ q^2\Sigma(q^2)}{q^2 + \Sigma^2(q^2)}
$$
\begin{equation}
\left. = \frac{1}{3\pi \alpha} \left( (p^2)^2 \frac{d\Sigma}{dp^2} \right)
\right|_{p^2= \Lambda^2} . \label{45}
\end{equation}

As was mentioned above, Eqs. (\ref{40}) and (\ref{41}) 
are written in the Landau gauge.  This gauge is
preferable in the rainbow approximation from the viewpoint of
satisfying the  Ward-Takahashi identities \cite{10}.  The transformation
to other gauges changes the vertex, i.e. it leads to an
approximation beyond the rainbow one.  However, the main results of
the analysis remain essentially the same \cite{18}.

The function $\Sigma(p^2)$, satisfying Eqs. (\ref{42}) - (\ref{44}), 
has the following ultraviolet asymptotics
(at $ \left. p^2 >> \Sigma_0^2 \equiv\Sigma^2(p^2)\right|_{p^2 = 0}$):
\begin{equation}
\Sigma(p^2) \rightarrow \tilde{A} \frac{\Sigma_0^2}{(p^2)^{1/2}}
\frac{1}{\omega} \sinh \left[ \omega \left( \frac{1}{2} \ln
\frac{p^2}{\Sigma_0^2} + \delta \right) \right] \label{46}
\end{equation}
at $\alpha < \alpha_c = \frac{\pi}{3} \: \: \left( \omega = (1 -
\alpha/\alpha_c)^{1/2} \right)$, 
\begin{equation}
\Sigma(p^2) \rightarrow \tilde{A} \frac{\Sigma_0^2}{(p^2)^{1/2}} \left(
\frac{1}{2} \ln \frac{p^2}{\Sigma_0^2} + \delta \right) \label{47}
\end{equation}
at $\alpha = \alpha_c$, and
\begin{equation}
\Sigma(p^2) \rightarrow \tilde{A} \frac{\Sigma_0^2}{(p^2)^{1/2}}
\frac{1}{\tilde{\omega}} \sin \left[ \tilde{\omega} \left( \frac{1}{2}
\ln \frac{p^2}{\Sigma_0^2} + \delta \right) \right] \label{48}
\end{equation}
at $\alpha > \alpha_c \left( \tilde{\omega} = \left(
\frac{\alpha}{\alpha_c} - 1 \right)^{1/2} \right)$.  
Here $\tilde{A}(\alpha)$ and $\delta(\alpha)$
are known functions of $\alpha$ . \footnote{
Note that in
the so-called linearized approximation \cite{12} for the SD equation (which
is a good one), the parameters $\tilde{A}$ and $\delta$ are
$$\tilde{A}(\alpha) = 2 \left[
\frac{\Gamma(1+\omega)\Gamma(1-\omega)}{\Gamma \left(
\frac{3+\omega}{2} \right) \Gamma \left( \frac{3-\omega}{2} \right)
\Gamma \left( \frac{1+\omega}{2} \right) \Gamma \left(
\frac{1-\omega}{2} \right) } \right]^{1/2}
$$
and
$$
\delta(\alpha) = \frac{1}{2\omega} \ln \left[ \frac{
\Gamma(1+\omega)}{\Gamma(1-\omega)} \cdot \frac{\Gamma
\left(\frac{3-\omega}{2} \right) \Gamma \left( \frac{1-\omega}{2}
\right)}{\Gamma \left( \frac{3+\omega}{2} \right) \Gamma \left(
\frac{1+\omega}{2} \right) } \right]
$$
at $\alpha < \alpha_c$.  At $\alpha > a_c$, $\omega$ is replaced by
$i \tilde{\omega}$ in these expressions.
}

We will calculate the effective potential along the critical line. 
Our aim is to show that while along the part of the critical line
(\ref{30}), with $g > \frac{1}{4}, \: \alpha < \alpha_c$, the potential has
the conventional, $\sigma$-model like, form, its form is rather
unusual along the critical line with $\alpha = \alpha_c$, where the
CPT occurs.

We begin by considering the potential along the part of the critical
line (\ref{30}).  From Eqs. (\ref{45}) and (\ref{46}), 
we find the chiral condensate in this case:
\begin{equation}
< 0 \vert \bar\psi \psi \vert 0 >_{J_1} = \frac{\tilde{A} \Sigma_0^2
\Lambda}{12 \pi \alpha \omega} \left[ e^{\omega \delta} (\omega - 1)
\left( \frac{\Lambda}{\Sigma_0} \right)^\omega + e^{-\omega \delta}
(\omega+1) \left( \frac{\Sigma_0}{\Lambda} \right)^\omega \right]. 
\label{49}
\end{equation}

The function $\Sigma_0(J_1)$ is determined from Eq. (\ref{43}) with $m^{(0)}
= -J_1$.  Substituting the condensate (\ref{49}) in equation (\ref{38}), 
we find:
$$
w(J_1) = - \frac{\tilde{A}^2 \Sigma_0^4}{96 \pi \alpha \omega^2} \left[
(\omega-1) Q^+ e^{2\omega\delta} \left(
\frac{\Lambda}{\Sigma_0}\right)^{2\omega}\right.
$$
\begin{equation}
\left. + (\omega+1) Q^-
e^{-2\omega\delta} \left(\frac{\Sigma_0}{\Lambda} \right)^{2\omega} +
2(1 - 4g) \right] \label{50}
\end{equation}
where
$$
Q^+ = \omega + 1 - \frac{4 g}{\omega+1}
$$
\begin{equation}
Q^- = \frac{4g}{1-\omega}- (1 - \omega) \label{51}
\end{equation}
and $\Sigma_0$ has to be considered here as a function of $J_1$. 
Realizing the Legendre transform of $w(J_1)$, we find the
effective potential:
$$
\tilde{V}(\rho_1) = J_1\rho_1 - w 
$$
$$= \frac{\tilde{A}^2 \Sigma_0^4}{96 \pi \alpha \omega^2} \left[ \left(1 -
\omega^2 - 4g \frac{1 - \omega}{1 + \omega} \right) e^{2 \omega \delta}
\left( \frac{ \Lambda}{\Sigma_0} \right)^{2 \omega} \right.
$$
\begin{equation}
\left. + \left( 1 - \omega^2 - 4g \frac{1+\omega}{1-\omega} \right)
e^{-2\omega\delta} \left( \frac{\Sigma_0}{\Lambda} \right)^{2\omega} +
2 \left( 4g - 1 - 2\omega^2 \right) \right], \label{52}
\end{equation}
where now $\Sigma_0$ has to be considered as a function of the
condensate
\begin{equation}
\rho_1 = < 0 \vert \bar\psi \psi \vert 0 >_{J_1} \simeq
\frac{\tilde{A} (\omega-1)}{12 \pi \alpha \omega} e^{\omega\delta}
\Lambda^{(\omega+1)} \Sigma_0^{2-\omega}. \label{53}
\end{equation}

A solution $\Sigma_0 = \bar{\Sigma}_0$ to the gap equation
$\frac{dV}{d\Sigma}_0 = 0$ (or $\frac{dV}{d\rho_1} = 0$) defines the
dynamical mass $m_{dyn} \equiv \bar\Sigma_0$ as a function of
$\Lambda, \alpha$ and  g:
\begin{equation}
m_{dyn}^2 \equiv \bar{\Sigma}_0^2 =\Lambda^2 e^{2\delta} \left[
 \left( \frac{1-\omega}{1+\omega} \right) \frac{g - \left(
\frac{\omega+1}{2}\right)^2}{g - \left( \frac{1-\omega}{2} \right)^2} 
\right]^{1/\omega}. \label{54}
\end{equation}
The limit $m_{dyn} \rightarrow 0$ defines the critical line (\ref{30}):  
$g - \frac{(\omega+1)^2}{4} = 0$.

Let us show that the potential (\ref{52}) can be rewritten as a
conventional, $\sigma$-model like, potential.

It is convenient to define the renormalized fields $(\bar\psi
\psi)_\mu$ and $(\bar\psi \gamma_5 \psi)_\mu$ as
$$
(\bar\psi \psi)_\mu = Z_m^{(\mu)} (\bar\psi \psi),
$$
\begin{equation}
(\bar\psi i \gamma_5 \psi)_\mu = Z_m^{(\mu)} (\bar\psi i \gamma_5
\psi) \label{55}
\end{equation}
where the renormalization constant is
\begin{equation}
Z_m^{(\mu)} = \frac{12 \pi \alpha \omega e^{-\omega
\delta}}{\tilde{A} (\omega -1)} \Lambda^{-(\omega -1)} \mu^{\omega-1} 
\label{56}
\end{equation}
(see Eq. (\ref{53}); notice that the renormalized composite fields
$(\bar\psi \psi)_\mu$ and $(\bar\psi i \gamma_5 \psi)_\mu$ are
defined in such a way as to have canonical dimension equal to 1).  Then
we find that
\begin{equation}
\sigma^{(\mu)} \equiv < 0 \vert ( \bar\psi \psi)_\mu \vert 0 > = \mu
\left( \frac{\Sigma_0}{\mu} \right)^{2-\omega} \label{57}
\end{equation}
at $\mu << \Lambda$ and $\Sigma_0 << \Lambda$.

We will express the potential (\ref{52}) through the chiral invariant
$\rho^{(\mu)} = \left( (\sigma^{(\mu)})^2 + (\pi^{(\mu)})^2
\right)^{1/2}$ (where $\pi^{(\mu)} \equiv < 0 \vert (\bar\psi i
\gamma_5 \psi)_\mu \vert 0 >$) and the mass parameter
\begin{equation}
\left. M_\mu^{(2)} = \frac{d^2 \tilde{V}}{d\rho^{(\mu)2}} 
\right|_{\rho^{(\mu)} = 0}. \label{58}
\end{equation}
It is easy to check that, along the critical line (\ref{30}),
the parameter $M_\mu^{(2)}$ is
\begin{equation}
M^{(2)}_\mu = \frac{\tilde{A}^2\mu^2}{48\pi\alpha\omega^2} 
\left(\frac{1-\omega}{1+\omega}\right) 
\left( (1+\omega)^2-4g\right)
e^{2\omega\delta}\left(\frac{\Lambda}{\mu}\right)^{2\omega}. \label{59}
\end{equation}
One can see that $M_\mu^{(2)} > 0$ and $M_\mu^{(2)} < 0$ from the
side of the symmetric and non-symmetric phases, respectively.  As
$\Lambda \rightarrow \infty$, with $M_\mu^{(2)}$ being fixed, the
potential (\ref{52}) is
\begin{equation}
\tilde{V} = \frac{1}{2} M_\mu^{(2)} (\rho^{(\mu)})^2 +
\frac{(2-\omega)\tilde{A}^2}{16 \pi^2 \omega(1-\omega^2)} \mu^4
\left( \frac{\rho^{(\mu)}}{\mu} \right)^{4/(2-\omega)}. \label{60}
\end{equation}
Thus, as was promised, we derived a $\sigma$-model like potential: 
the sign of $M_\mu^{(2)}$ defines two different phases, and the value
$M_\mu^{(2)} = 0$ corresponds to the critical line with $0 < \alpha <
\alpha_c = \pi/3$.  Notice that the parameter $M_\mu^{(2)}$ appears 
in $\tilde{V}$ as a
result of the dynamical transmutation:  in the continuum limit,
$\Lambda \rightarrow \infty$, the dimensional parameter $M_\mu^{(2)}$
replaces the dimensionless coupling constant $g$.

Note also that, because the anomalous dimension $\gamma_m$ of
$\bar\psi \psi$ and $\bar\psi i \gamma_5 \psi$ is $\gamma_m = 1 +
\omega$ (see Eq. (\ref{32})), the dynamical dimension of these operators,
and therefore of $\rho^{(\mu)}$, is $d_{\bar\psi \psi} = 3 - \gamma_m
= 2 - \omega$. Therefore the second term in the potential (\ref{60}) has
the dynamical dimension equal to 4, i.e. it preserves the conformal
symmetry.  The first, massive, term in the potential breaks 
this symmetry.

One can check that, taking $\mu = m_{dyn} \equiv \bar\Sigma_0$, the potential
in the non-symmetric phase can be rewritten as
\begin{equation}
\tilde{V} = \frac{\tilde{A}}{16 \pi^2 \omega(1-\omega^2)} 
\bar\Sigma_0^4 \left[
(2-\omega) \left( \frac{\rho_r}{\bar\Sigma_0} \right)^{4/(2-\omega)} -
2 \left( \frac{\rho_r}{\bar\Sigma_0} \right)^2 \right], \label{61}
\end{equation}
where $\rho_r \equiv \rho^{(\mu)} \vert_{\mu = \bar\Sigma_0}$.

As shown in Refs. \cite{4,17}, the kinetic term and terms with a larger
number of derivatives in the effective action
are also finite in the continuum
limit. \footnote{
Actually, as was shown in Ref. \cite{17}, all
these terms are conformal invariant
}
A heuristic explanation of this fact is simple:  the most
severe ultraviolet divergences always occur in the effective
potential;  therefore the finiteness of the potential implies the
finiteness of other terms in the action.  This implies that around
the critical line (\ref{30}), both in the symmetric and non-symmetric phase,
the light excitations include, besides fermions and photon, composite
$\sigma$ and $\pi$ particles.

Let us now turn to the part of the critical line with $\alpha =
\alpha_c$, where the CPT takes place, and show that the character
of the phase transition is essentially different there.

We begin by considering the critical point $(\alpha, g) = (\alpha_c,
\frac{1}{4})$.  Calculating the potential in the same way as the potential
(\ref{52}) for $\alpha < \alpha_c$, we find that at $\alpha = \alpha_c$,
around $g = \frac{1}{4}$, it is
\begin{equation}
\tilde{V} = - \frac{\tilde{A}^2 \Sigma_0^4}{8 \pi^2} \left[ (4g-1) L^2 - 8gL
+ 4g + \frac{3}{2} \right] \label{62}
\end{equation}
where $L=\ln \left( \Lambda e^\delta / \Sigma_0 \right)$.  The gap
equation $\frac{d \tilde{V}}{d\Sigma_0} = 0$ 
yields the dynamical mass at $g > \frac{1}{4}$:
\begin{equation}
m_{dyn}^2 \equiv \bar\Sigma_0^2 =
\Lambda^2 e^{2(\delta-1)} \exp \left( -
\frac{1}{(g - \frac{1}{4})} \right). \label{63}
\end{equation}

At $\alpha = \alpha_c$, the relation between $\rho_r = \rho^{(\mu)}
\vert_{\mu = \bar\Sigma_0}$ and $\Sigma_0$ is
\begin{equation}
\rho_r = \bar\Sigma_0 \left( \frac{\Sigma_0}{\bar\Sigma_0} \right)^2 
\label{64}
\end{equation}
(compare with Eq. (\ref{57}); in this case
the renormalization constant is
$Z_m^{(\mu)}
= \frac{- 2 \pi^2}{\tilde{A} \Lambda \mu} \left( \ln
\frac{\Lambda e^\delta}{\mu} - 1 \right)^{-1}$).  Taking the
continuum limit $\Lambda \rightarrow \infty$ in Eq. (\ref{62}) (with
$\bar\Sigma_0$ being fixed) we come to the expression for the
potential in the non-symmetric phase:
\begin{equation}
\tilde{V} = \frac{\tilde{A}^2}{16 \pi^2} \bar\Sigma_0^2 \rho_r^2
\left( \ln
\frac{\rho_r^2}{\bar\Sigma_0^2} - 1 \right). \label{65}
\end{equation}
But what is the form of the potential (in the continuum limit) from
the side of the symmetric phase? Let us show that there the potential
$\tilde{V}$ goes to infinity as $\Lambda \rightarrow \infty$.
\footnote{
There was an attempt to make the effective potential finite at
$\alpha = \alpha_c$ in the $\Lambda \rightarrow \infty$ limit
\cite {7}. This may define a different continuum theory
than that
discussed in this paper.
}
Indeed Eq. (\ref{62}) implies that to get a finite $\tilde{V}$ as $\Lambda
\rightarrow \infty$, the coupling constant $g$ must have the behavior
\begin{equation}
4g - 1 \sim \frac{2}{L} > 0, \label{66}
\end{equation}
i.e. it {\it must} go to the critical value $g_c = \frac{1}{4}$ from
the side of the non-symmetric phase.  In the symmetric phase the
potential $\tilde{V} \rightarrow \infty$ as $\Lambda \rightarrow \infty$.

What is the physical meaning of this result?  As in the case of 
the Gross-Neveu model, it implies that there are no composite 
light resonances in the symmetric phase at $\alpha = \alpha_c, \; 
g < g_c = \frac{1}{4}$. 
Indeed, the relations (\ref{8}) and (\ref{63}) imply that in the symmetric 
phase 
\begin{equation}
|M_{\pi}^2 | = | M_{\sigma}^2 | \sim \Lambda^2 \exp \left( 
\frac{1}{(\frac{1}{4} - g) } \right) > \Lambda^2. \label{67}
\end{equation}

This conclusion correlates with the point that, unlike the case with
the critical line with $\alpha < \alpha_c$, the parameter $M_\mu^{(2)}$ is
not well defined at $(\alpha, g) = \left( \alpha_c, \frac{1}{4}
\right)$:  it is $M_\mu^{(2)} \rightarrow +\infty$ and $M_\mu^{(2)}
\rightarrow -\infty$ in the symmetric and non-symmetric phase,
respectively (see Eq. (\ref{59})).

Thus the CPT takes place at $(\alpha, g ) = (\alpha_c , \frac{1}{4})$. 

Let us show that a similar situation takes place along the whole
critical line with $\alpha = \alpha_c$, where the CPT takes place.

Around the critical line with $\alpha = \alpha_c$, the condensate $<
0 \vert \bar\psi \psi \vert 0 >_J$ is
\begin{equation}
< 0 \vert \bar\psi \psi \vert 0 >_J = -\frac{\tilde{A} \Sigma_0^2
\Lambda}{2 \pi^2} \left( \frac{ \sinh \theta}{\omega} - \cosh \theta
\right) \label{68}
\end{equation}
at $\alpha < \alpha_c$, (the symmetric phase), and
\begin{equation}
< 0 \vert \bar\psi \psi \vert 0 >_J = - \frac{\tilde{A} \Sigma_0^2
\Lambda}
{2 \pi^2} \left( \frac{\sin \tilde{\theta}}{\tilde{\omega}} -
\cos
\tilde\theta \right) \label{69}
\end{equation}
at $\alpha > \alpha_c$ (the non-symmetric phase).  Here $\theta =
\omega \ln \left( \Lambda e^\delta / \Sigma_0 \right)$ and
$\tilde{\theta} = \tilde{\omega} \ln \left( \Lambda e^\delta /
\Sigma_0 \right)$, and Eqs. (\ref{45}), (\ref{46}) and (\ref{48}) 
were used in the derivation of expressions (\ref{68}) and (\ref{69}).

Now the effective potential can be derived in the same way as
before.  It is
$$ \tilde{V} = \frac{ \tilde{A}^2 \Sigma_0^4}{96 \pi \alpha \omega^2} 
\left[
\left(1 - \omega^2 - 4g \frac{1-\omega}{1+\omega} \right) \left(
\frac{\Lambda e^\delta}{\Sigma_0} \right)^{2\omega} +
\left( 1 - \omega^2 - 4g \frac{1+\omega}{1-\omega} \right)
\left(\frac{\Sigma_0}{\Lambda e^\delta} \right)^{2 \omega} \right.
$$
\begin{equation}
\left. + 4g - 1 - 2\omega^2 \right] \simeq - \frac{\tilde{A}^2
\Sigma_0^4}{24 \pi \alpha} \left[ \left( 4g-1 \right) L^2 - 8gL + 4g
+ \frac{3}{2} \right] \label{70}
\end{equation}
at $\alpha < \alpha_c$ and 
$\omega << 1$ $(g < \frac{1}{4})$, and
$$
\tilde{V} = \frac{ \tilde{A}^2 \Sigma_0^4}{48 \pi \alpha \tilde{\omega}^2}
\left[ 1 - 4g - 2 \tilde{\omega}^2 - \left( 1+ \tilde{\omega}^2
\right) \cos \left( 2 \tilde{\omega}L \right) \right.
$$
\begin{equation}
\left. + 4g \cos \left( 2 \tilde{\omega} L - 2 \arctan \tilde{\omega}
\right) \right] \label{71}
\end{equation}
at $\alpha >  \alpha_c$ and 
$\tilde{\omega} << 1$ $(g < \frac{1}{4})$. 

Notice that at $\alpha = \alpha_c = \pi/3$, the expression (\ref{70}) for
$\tilde{V}$ coincides with the expression (\ref{62}).  
Therefore we conclude that
in the symmetric phase, along the whole critical line with $\alpha =
\alpha_c$, the collective $\sigma$ and $\pi$ excitations decouple
from the infrared dynamics.  In this phase, the conformal dynamics of
massless fermions and photons is realized.

Let us now consider the potential (\ref{71}) in the non-symmetric phase. 
The gap equation $ \frac{ d \tilde{V}}{d \Sigma_0} = 0$ yields the following
solutions for $\Sigma_0$ at $(1-4 g ) >> \tilde{\omega} << 1$:
\begin{equation}
\tilde{\omega} \bar L = \pi n - \tilde{\omega} \frac{1+4g}{1-4g}\; ;
\; n = 1, 2,... \label{72}
\end{equation}
\begin{equation}
\tilde{\omega} \bar L = \pi n + \frac{3}{2} \tilde{\omega} \; ; \; n
= 1,2,... \label{73}
\end{equation}
where $\bar L = \ln \frac{\Lambda e^\delta}{\bar\Sigma_0}$.  One can
check that while all the solutions (\ref{72}) correspond to minima of 
$\tilde{V}$,
the solutions (\ref{73}) correspond to maxima of the potential.  Actually,
only the global minimum, corresponding to $n=1$, defines the stable
vacuum.  Therefore the dynamical mass is
\begin{equation}
m_{dyn}^2 \equiv \bar\Sigma_0^2 = \Lambda^2 \exp \left( 2 \delta +
\frac{ 2(1+4g)}{1 - 4g} \right) \exp \left( - \frac{2 \pi}{ \left(
\frac{\alpha}{\alpha_c} - 1 \right)^{1/2}} \right). \label{74}
\end{equation}
The expression (\ref{74}) is valid at $(1-4 g ) >> \tilde{\omega} <<1$. 
At $(1-4 g ) << \tilde{\omega} <<1$, we find from Eq. (\ref{71}) that 
\begin{equation}
m_{dyn}^2 = \Lambda^2 \exp (2\delta -3) \exp \left( 
- \frac{\pi}{(\frac{\alpha}{\alpha_c} - 1 )^{1/2}} \right). \label{75}
\end{equation}
In the continuum limit $\Lambda \rightarrow \infty$,
with $\bar\Sigma_0$
being fixed (i.e., lim $\tilde\omega \bar L \rightarrow \pi$
as $\Lambda
\rightarrow \infty$ and $\tilde{\omega} \rightarrow 0$), 
we find the expression for $\tilde{V}$:
\begin{equation}
\tilde{V} = \frac{ \tilde{A}^2 \Sigma_0^4 (\rho_r)}{16 \pi^2} \left[ 8
\left( \frac{1}{4} - g \right) \ln^2 \frac{\Sigma_0 (\rho
_r)}{\bar\Sigma_0} + 4\ln \frac{ \Sigma_0 (\rho_r)}{\bar\Sigma_0} - 1 \right].
 \label{76}
\end{equation}
Here $\Sigma_0$ is considered as a function of a renormalized field
$\rho_r = (\sigma_r^2 + \pi_r^2)$ which is defined in the following
way.  One finds from Eq. (\ref{69}) that, in the continuum limit, the
condensate is:
\begin{equation}
< 0 \vert \bar\psi \psi \vert 0 >_J \simeq \frac{ \tilde{A}
\Sigma_0^2 \Lambda}{\pi^2 (4g - 1)} \left( 1 + \frac{1 - 4g}{2} \ln
\frac{ \Sigma_0}{\bar\Sigma_0} \right). \label{77}
\end{equation}
Therefore it is appropriate to define the renormalized fields
$<\bar\psi \psi >_r$ and $< \bar\psi i \gamma_5\psi >_r$ with the
renormalization constant
\begin{equation}
Z_m = \frac{ \pi^2 (4g - 1)}{\tilde{A}} \Lambda^{-1}
\bar\Sigma_0^{-1}. \label{78}
\end{equation}
Then one finds that
\begin{equation}
\sigma_r = Z_m < 0 \vert \bar\psi \psi \vert 0 >_J = \Sigma_0^2
\bar\Sigma_0^{-1} \left[ 1 + \frac{ 1 - 4g}{2} \ln \frac{
\Sigma_0}{\bar\Sigma_0} \right]. \label{79}
\end{equation}
The function $\bar\Sigma_0 (\rho_r)$ in Eq. (\ref{76}) is defined from Eq.
(\ref{79}), with $\sigma_r$ replaced by $\rho_r$.

The potential (\ref{76}) describes composite $\sigma$ and $\pi$ particles
in the non-symmetric phase.

Thus there is an abrupt change of the number of light excitations and
the character of the dynamics, as the critical line is crossed, along 
the whole critical line with $\alpha = \alpha_c$, where
the CPT takes place.  

It is instructive to compare the effective action (generating
functional for proper vertices of the operators $\bar\psi \psi$ and
$\bar\psi i \gamma_5 \psi$)  considered here with the 
effective action $\Gamma$ for auxiliary fields, 
whose derivation in the gauged NJL model
was considered in Refs. \cite{16,17,7}.  The action $\Gamma(\sigma, \pi)$
is derived by rewriting the Lagrangian density (\ref{29}) as
\begin{equation}
{\cal L} = - \frac{1}{4} \left( F_{\mu \nu} \right)^2 + \frac{1}{2}
\left[ \bar\psi, (i \gamma^\mu D_\mu) \psi \right] - \bar\psi (\sigma
+ i \gamma_5 \pi) \psi - \frac{1}{2 G} (\sigma^2 + \pi^2), \label{80}
\end{equation}
and then integrating out both the fermion and photon fields.  In the
mean field approximation, the path integral over $\sigma$ and $\pi$
is dominated by the stationary points of $\Gamma$:  $\frac{\delta
\Gamma}{\delta \pi} = \frac{\delta \Gamma}{\delta \sigma} = 0$.

It is clear, however, that the trick of introducing the fields
$\sigma$ and $\pi$ in Eq. (\ref{80}) cannot be applied to pure QED $(G
= 0)$.  It is also unclear how well is the mean field approximation
around the part of the critical line with $\alpha = \alpha_c$, where
the QED dynamics dominate.

Comparing the expression for the potential $\tilde{V}$ with that for the
potential $V$, derived in Refs. \cite{16,17,7}, one finds that while they
coincide around the part (\ref{30}) of the critical line, with $\alpha <
\alpha_c$ and $g > \frac{1}{4}$ (where the four-fermion interactions
dominate), they are different around the critical line with $\alpha =
\alpha_c$, where the CPT takes place.  In  particular, the
counterpart of the expressions (\ref{62}) and (\ref{76}) 
for $\tilde{V}$ are \cite{16,7} 
\begin{equation}
V = - \frac{ \tilde{A}^2 \Sigma_0^4}{8 \pi^2} \left[ 
\left( \frac{4g-1}{4g} \right) L^2 -2L + \frac{5}{2} \right], \label{81}
\end{equation}
and \cite{17}
\begin{equation}
V = \frac{ \tilde{A}^2 \Sigma_0^4}{16 \pi^2} \left[ \frac{2}{g}
\left( \frac{1}{4} - g \right) \ln^2 \frac{\Sigma_0}{\bar\Sigma_0} +
4 \ln \frac{\Sigma_0}{\bar\Sigma_0} - 1 \right], \label{82}
\end{equation}
respectively. 
Unlike the expressions (\ref{62}) and (\ref{76}) for $\tilde{V}$ , 
$V$ is singular at $g = 0$ (pure QED). 

This feature of $V$ reflects the point that at $\alpha
\simeq \alpha_c$, $g < \frac{1}{4}$, when QED interactions dominate,
the mean field approximation is not good enough and one has to
consider quantum fluctuations of the fields $\sigma$ and $\pi$.  On
the other hand, the generating functional $\tilde{\Gamma}$ adequately
describes the dynamics along the whole critical line.

A similarity between the dynamics in quenched QED4 and 
$D$-dimensional GN model considered in Sec. 3 is evident. 
At $\alpha < \alpha_c = \frac{\pi}{3}$ $ (D>2)$ a $\sigma$-model like 
phase transition is realized in quenched QED4 (GN model); 
at $\alpha = \alpha_c$ $(D=2)$ the CPT takes place in these models. 
However, there is an essential difference between the CPT phase 
transitions in these two models. While in the GN model, 
the symmetric phase, with $g<0$, is infrared free, the symmetric phase in 
quenched QED is a Coulomb phase, describing interactions between massless 
fermions and photons. 

As was indicated in Sec. 3, a marginal operator is responsible for the 
breakdown of the conformal symmetry in the non-symmetric phase in the 2-dimensional GN model (see Eq. (\ref{27})). This leads to an essential singularity 
in the expression for the order parameter $\bar{\rho}$ (\ref{18}). 
This in turn cures fine tuning problem which takes place at $D>2$, 
where relevant (superrenormalized) operators break the conformal symmetry. 

A similar situation takes place in quenched QED4. While at 
$\alpha < \alpha_c$, the (relevant) mass operator breaks the conformal 
symmetry (see Eq. (\ref{60})), at $\alpha = \alpha_c$, it is 
broken (in non-symmetric) phase by a marginal operator. 
Actually, as shown in the Appendix, the equation for the conformal 
anomaly has the following form at $(\alpha ,g)=(\alpha_c , \frac{1}{4})$: 
\begin{eqnarray}
\partial^{\mu} D_{\mu} &=& \theta_{\mu}^{\mu} = 
\lim_{\Lambda \to \infty, g \to \frac{1}{4}} \frac{G}{2} 
\frac{\beta (g)}{g} 
\left[ (\bar{\psi}\psi)^2 + (\bar{\psi} i \gamma_5 \psi)^2 \right] 
\nonumber \\ 
&=&  \lim_{\Lambda \to \infty, \alpha \to \alpha_c} \lim 
\frac{\beta (\alpha)}{4\alpha} F_{\mu \nu}F^{\mu \nu},  \label{83}
\end{eqnarray}
where $\beta (g) = \frac{\partial g}{\partial \ln \Lambda} = 
-2 (g - \frac{1}{4})^2$ is determined from Eq. (\ref{63}), and 
$\beta (\alpha) = \frac{\partial \alpha}{\partial \ln \Lambda} = 
-\frac{4}{3} (\frac{\alpha}{\alpha_c} -1)^{3/2}$ is determined 
from Eq. (\ref{75}). 
As was pointed above, at $(\alpha , g) = (\alpha_c , \frac{1}{4})$ 
the dynamical dimension of the operator 
$\left[ (\bar{\psi}\psi)^2 + (\bar{\psi} i \gamma_5 \psi)^2 \right] $ 
is $d=4$ and, therefore, it is indeed a marginal operator 
(the operator $F_{\mu \nu}F^{\mu \nu}$ in Eq. (\ref{83}) is also of 
course marginal). 

At $\alpha = \alpha_c$ and $g <\frac{1}{4}$, the equation for 
the conformal anomaly is (see the Appendix) 
\begin{equation}
\partial^{\mu} D_{\mu} = \theta_{\mu}^{\mu} = 
\lim_{\Lambda \to \infty, \alpha \to \alpha_c} \frac{\beta (\alpha)}{4\alpha} 
F_{\mu\nu}F^{\mu\nu} \label{84}
\end{equation}
where $\beta (\alpha) = 
-\frac{2}{3} (\frac{\alpha}{\alpha_c} -1)^{3/2}$ is determined from 
Eq. (\ref{74}). Here, again, 
$\partial^{\mu}D_{\mu}$ is a marginal operator. 
In the next section, we shall summarize the main features of the CPT. 
We shall also discuss the phase transition in QED3. 

\section{General Features of the CPT. A pseudo-CPT in QED3. }

Now we are ready to summarize the main features of the CPT. 

There is an abrupt change of the spectrum of light excitations, 
as the critical point $z = z_c$ is crossed, in the CPT. 
As was shown in Sec. 2, this property is general and reflects the presence 
of an essential singularity at $z=z_c$ in the scaling function $f(z)$. 

The CPT is (though continuous) a non-$\sigma$-model like phase transition. 
This implies a specific form of the effective action, in particular, 
the effective potential, for the light excitation near $z=z_c$. 
While the potential does not exist in the continuum limit in the symmetric 
phase, it has infrared singularities at $\rho = 0$ in the non-symmetric 
phase ($\rho $ is a generic notation for fields describing the light 
excitations). 
As a result, unlike the $\sigma$-model like phase transition, one cannot 
introduce parameters $M^{(2n)} = \frac{d^{2n} V}{d \rho^{2n}}
\left. \right|_{\rho = 0}$ which would govern the phase transition:
all of them are equal either to zero or to infinity. 

The infrared singularities in the effective potential imply the presence 
of long range interactions as $\rho \to 0$. This is turn connected with 
an important role of the conformal symmetry in the CPT. 
In the examples considered in Sec. 3 and 4, while the symmetric phase is 
conformal invariant, there is a conformal anomaly in the non-symmetric phase: 
the conformal symmetry is broken by a marginal operator. 
The latter allows to get rid of the fine tuning problem in such a dynamics. 
We shall return to the problem of the effective action in the CPT 
in the next section. 

Because of the abrupt changing the spectrum of light excitations at 
$z=z_c$, the very notion of the universality class for the dynamics 
with the CPT seems rather delicate. 
For example, in both GN model and QCD, at the critical point 
$(g=0$ and $\alpha^{(0)}=0$, respectively), and at finite cutoff 
$\Lambda$, the theories are free and their infrared dynamics
are very different from the infrared dynamics 
in the non-symmetric phases of these theories (at $g>0$ and 
$\alpha^{(0)}>0$, respectively). 
This is a common feature of the CPT: 
around the critical point, the infrared dynamics in the symmetric 
and non-symmetric phases are very different. However, in the  
non-symmetric phase,   
the hypothesis of universality has to be applied to the region of 
momenta $p$ satisfying $\bar{\rho} << p << \Lambda$, where $\bar{\rho}$ 
is an order parameter. In that region, critical indices 
(anomalous dimensions) of both elementary and composite local operators 
in near-critical regions of symmetric and non-symmetric phases are nearly 
the same: 
the critical indices are continuous functions of $z$ around $z=z_c$
\footnote{
However, because of explicit conformal symmetry breaking in 
the non-symmetric phase, there are additional logarithmic factors 
(such as $(\ln \frac{p}{\bar{\rho}})^c$) in Green's functions in that phase. 
}. 
On the other hand, since the infrared
dynamics (with $p \sim \bar{\rho}$ and 
$p << \bar{\rho}$) abruptly changes as the critical 
point $z=z_c$ is crossed, the low energy effective actions in
the symmetric and non-symmetric phases are different.
 
If there is a
perturbative running of the coupling in the symmetric phase, it will
lead to perturbative violation of the conformal symmetry (see
Sec. 7). 
However, this does not change the most characteristic 
point of the CPT: 
the abrupt change of the spectrum of light excitations at 
$z=z_c$ discussed above. The reason is that there is an additional,
nonperturbative, source  of the breakdown of the conformal symmetry
in the non-symmetric phase, which provides the creation of light
composites.

The conception of the CPT, in a slightly modified form, can be also 
useful for a different type of dynamics. As an example, let us consider 
QED3 with massless four-component fermions \cite{20}. 
It is a superrenormalizable theory where ultraviolet dynamics plays 
rather a minor role. As was shown in Refs. \cite{21,22}, when the number of 
fermion flavors  $N_f$ is less than $N_{cr}$, with $3< N_{cr} < 4$, 
there is dynamical breakdown of the flavor $U(2N_f)$ symmetry in the model, 
and fermions acquire a dynamical mass
\footnote{
We are aware that there is still a controversy concerning this result: 
some authors argue that the generation of a fermion mass occurs at all 
values of $N_f$ \cite{23}. For a recent discussion supporting the 
relation (\ref{85}), see Ref. \cite{24}.
}:
\begin{equation}
m_{dyn} \sim \alpha_3 \exp \left[ -\frac{2\pi}{\sqrt{N_{cr} / N_f
-1}} \right], 
  \label{85}
\end{equation}
where the coupling constant
$\alpha_3 = e^2/4\pi$ is dimensional in QED3. 

Though this expression resembles the expression (\ref{74}) for 
the dynamical mass in quenched QED4, where $\Lambda$ plays the role of 
$\alpha_3$ and $\alpha$ plays the role of $N_f$, the phase transition at 
$N_f=N_{cr}$ is, strictly speaking, not the CPT. 
Indeed, because of superrenormalizability of QED3, the ultraviolet cutoff 
$\Lambda$ is irrelevant for the dynamics leading to relation (\ref{85}). 
Also, since $\alpha_3$ is dimensional, the conformal symmetry is broken 
in both symmetric $(N_f > N_{cr})$ and non-symmetric
$(N_f < N_{cr})$ phases. 

Nevertheless, the consideration of the spectrum of light 
(with $(M^2 << \alpha_3^2)$ ) 
excitations in this model can be done along the 
lines used in Sec. 2. 
In agreement with the result of Ref. \cite{6}, where the BS equation 
was used, one concludes that there are no light resonances 
(with $(M^2 << \alpha_3^2)$ ) in the symmetric phase of QED3 and 
that there is an abrupt change of the spectrum of light 
excitations at $N_f = N_{cr}$. 

It is appropriate to call the phase transition in QED3 a  
pseudo-CPT: in the non-symmetric phase, at $N_f < N_{cr}$, a new, 
nonperturbative, source of the breakdown of the conformal symmetry
occurs.

\section{The Effective Action in Theories with 
the CPT and the Dynamics of the Partially Conserved Dilatation
Current}

In this section we shall discuss the properties of the effective 
action in theories with the CPT in more detail. 
In particular we shall consider a connection of the dynamics of the CPT 
with the hypothesis of the partially conserved dilatation current 
(PCDC) \cite{25,26,27,41}.

The effective potentials derived in the 2-dimensional GN model
(see Eqs. (\ref{16})) and (\ref{24})) and in quenched QED4 
with $(\alpha , g) = (\alpha_c, \frac{1}{4})$ (see Eqs. (\ref{62}) 
and (\ref{65})) have a similar form. 

Moreover, one can show that the kinetic term and terms with 
higher number of derivatives in both the GN model and quenched QED4 
are conformal invariant \cite{8,17}. 
In other words, the conformal anomaly comes only from the effective 
potential in both these models. 

This point is intimately connected with the PCDC dynamics. 
In order to see this, let us determine the divergence of the dilatation current in these models. Eq. (\ref{24}) implies that 
\begin{equation}
\partial^{\mu} D_{\mu} = \theta^{\mu}_{\mu} = 
-\frac{2N_c}{\pi} \rho^2 \label{86}
\end{equation}
in the GN model, and Eq. (\ref{65}) yields
\begin{equation}
\partial^{\mu} D_{\mu} = \theta^{\mu}_{\mu} =  
- \frac{\tilde{A}^2}{4 \pi^2} m_{dyn}^2 \rho^2   \label{87}  
\end{equation}
in quenched QED4 with $\alpha = \alpha_c$, where the CPT takes place 
$(m_{dyn} \equiv \bar{\Sigma}_0)$. 
Now, recall that the dynamical dimension $d_{\rho}$ of the field $\rho$ is 
$d_{\rho}=1$ and $d_{\rho}=2$ in the GN model and in quenched QED4 (with 
$\alpha = \alpha_c$), respectively. 
Therefore Eqs. (\ref{86}) and (\ref{87}) assure that the dynamical 
dimension of the operator $\theta^{\mu}_{\mu}$ coincides with its 
canonical dimension: 
$d_{\theta}=2$ and $d_{\theta}=4$ in the 2-dimensional GN model
and quenched 
QED4, respectively. 
This implies the the realization of the PCDC hypothethis in these 
models  \cite{25,26,27,41}: the operator $\theta^{\mu}_{\mu}$
has the correct transformation properties under dilatation
transformations.

In the renormalization group language, this means that the conformal 
symmetry in these models is broken by marginal (renormalized) operators 
and not by relevant (superrenormalized) ones (irrelevant 
(nonrenormalized) operators contribute only
small corrections in the infrared 
dynamics). 

Though these two models are very special, one may expect that at least 
some features of this picture will survive in the general case of theories 
with the CPT. In particular, one may expect that in the general case 
the effective potential has the form 
\begin{equation}
V(\rho ) = C \bar{\rho}^D \left( 
\frac{\rho}{\bar{\rho}} \right)^{\frac{D}{d_{\rho}}} F( 
\ln \frac{\rho}{\bar{\rho}} ) \label{88}
\end{equation}
where $C$ is a dimensionless constant and $F(x)$ is a (presumably) 
smooth function. 

The contribution of $V(\rho)$ (\ref{88}) into the conformal anomaly 
is of the form 
\begin{equation}
\theta^{\mu}_{\mu} \sim \bar{\rho}^D \left( \frac{\rho}{\bar{\rho}} 
\right)^{\frac{D}{d_{\rho}}} 
F'(\ln \frac{\rho}{\bar{\rho}}),   \label{89}
\end{equation}
where $F'(x) = \frac{d F}{d x}$, i.e., in the general case, logarithmic 
factors may destroy the covariance (with respect to dilatation 
transformations) of the relation for the conformal anomaly. 
Actually this takes place already in quenched QED4 with $\alpha = 
\alpha_c$ but $g < \frac{1}{4}$. Indeed, as follows from 
Eqs. (\ref{76}) and (\ref{79}), the logarithmic factors occur
in the equation for 
the conformal anomaly in that case. 

Also, one should expect that the conformal invariance of the kinetic term 
and terms with higher number of derivatives may also be destroyed by logarithmic terms. 

It is clear that the effective action in theories with the CPT 
are very different from that in the 4-dimensional
linear $\sigma$-model and 
Nambu-Jona-Lasinio model, where the conformal symmetry is broken 
by relevant operators and the chiral phase transition is a mean-field one. 

This point can be relevant for the description of the low energy dynamics 
in QCD and in models of dynamical electroweak symmetry breaking. 
In particular, as was already pointed out in Ref. \cite{27}, the 
low energy dynamics are very sensitive to the value of the dynamical 
dimension $d_{\rho}$.

\section{Phase Diagram in a $SU(N_c)$ Gauge Theory}

In this section, we will consider the phase diagram with respect to 
the bare coupling constant $\alpha^{(0)}$ and the number of fermion flavors 
$N_f$ in a 4-dimensional $SU(N_c)$ vector like gauge theory \cite{2}. 
In particular, we will discuss a recent suggestion \cite{5} that 
the phase transition with respect to $N_f$ in that theory resembles the 
phase transition (with respect to the coupling constant) in
quenched QED4 at 
$\alpha = \alpha_c$. 

A starting point of the analysis of Refs. \cite{2} and \cite{5} is 
the presence of an infrared fixed point in the two-loops $\beta$ 
function of a $SU(N_c)$ theory, when the number of fermion flavors 
$N_f$ is large enough. Recall that the perturbative $\beta$ function 
in that theory is 
\begin{equation}
\beta(\alpha) = - b \alpha^2 -c \alpha^3 - d \alpha^4 - \cdot \cdot \; . 
  \label{90}  
\end{equation}
In the case of the $N_f$ fermions in the fundamental representation, 
the first two coefficients are \cite{28}:
\begin{eqnarray}
b &=& \frac{1}{6\pi} ( 11 N_c - 2 N_f), \nonumber \\
c &=& \frac{1}{24\pi^2} ( 34 N_c^2 - 10 N_c N_f - 3 \frac{N_c^2 -1}{N_c} N_f).
   \label{91}  
\end{eqnarray}
While these two coefficients are invariant under change of a renormalization 
scheme, the higher-order coefficients are scheme-dependent. 
Actually, there is a renormalization scheme in which all the higher-order 
coefficients vanish \cite{29}. Therefore there is at least one 
renormalization scheme in which the two-loop $\beta$ function 
is (perturbatively) exact. We will use such a renormalization scheme. 

The theory is asymptotically free if $b>0$ $(N_f < N_f^{**} \equiv 
\frac{11}{2}N_c)$. 
If $b>0$ and $c<0$, the $\beta$ function has a zero, 
corresponding to a infrared-stable fixed point, at
\begin{equation}
\alpha = \alpha^*  = - \frac{b}{c}.  \label{92}  
\end{equation}
When $N_f$ is close to $N_f^{**} = \frac{11}{2}N_c$, the ratio 
$|\frac{b}{c}|$, and therefore the value of $\alpha^*$, is small. 
The value of the fixed point $\alpha^*$ increases with decreasing $N_f$, 
and this fixed point disappears at the value $N_f= N_f^{*}$, 
when the coefficient $c$ becomes positive ($N_f^*$ is $N_f^* \simeq 8.05 $ 
for $N_c = 3$). 

It is convenient to consider $N_f$ as a continuous parameter and to 
study the dynamics as $N_f$ is varied. Note that since $N_f$ 
appears analytically in the path integral of the theory, one can give 
a non-perturbative meaning to the theory with non-integer $N_f$. 

Unlike ultraviolet-stable fixed points, defining dynamics at high momenta, 
infrared-stable fixed points (defining dynamics at low momenta) are very 
sensitive to nonperturbative dynamics leading to the generation of particle
masses. For example, if fermions acquire  
a dynamical mass, they decouple from the infrared dynamics. 
Then, only gluons will contribute to the $\beta$ function, and as a result, 
the perturbative infrared-stable fixed point in the $\beta$ function will 
disappear. 

Thus the crucial question is the interplay between the value $\alpha^*$ 
of the infrared-stable fixed point and the chiral dynamics. 

In Fig. 2, the phase diagram suggested in Ref. \cite{2} is shown. 
The authors considered the hamiltonian lattice gauge theories with 
Kogut-Susskind fermions. The main features of this diagram are the following. 
The vertical line (with $N_f < N_f^* < N_f^{**} = \frac{11N_c}{2}$) 
corresponds to a first-order phase transition, dividing a weak-coupling 
phase, possessing a continuum limit $\Lambda \to \infty$, from 
a strong-coupling phase. Notice that there is spontaneous chiral symmetry 
breaking in both these phases: 
the chiral order parameter jumps without 
vanishing at this phase transition: 

Note the following points \cite{2}:
 
a) The vertical line occurs because the chiral $\beta$ function is positive 
for large $g^{(0)}$ and negative for small $g^{(0)}$, where it 
coincides with the perturbative $\beta$ function. Thus $\beta$ function has 
an infrared-stable zero at some intermediate $g^{(0)}$. 
The line of these zero is the vertical line at which the first-order 
phase transition occurs (it cannot be second-order because $\beta$ 
has no ultraviolet-stable zeroes); 

b) The number $N_f$ refers to the number of fermion fields in 
the formal continuum theory which is twice the number of single component 
lattice fields. Therefore the minimal (non-zero) value of $N_f$ is 
$N_f = 2$. 
The vertical line ends at this value since in pure gluodynamics there is 
apparently no phase transition between weak-coupling and strong-coupling 
phases \cite{30}. 

The right-hand portion of the curve on the diagram, 
separating symmetric and non-symmetric phases occurs due to the following 
reason. At large enough values of the coupling, spontaneous chiral symmetry 
breaking takes place for any number $N_f$ of Kogut-Susskind fermions. 
Then the authors of Ref. \cite{2} argue that it is not reasonable to 
allow spontaneous chiral symmetry breaking to persist bellow some finite 
$g^{(0)}_c (N_f)$. As a result they suggest the existence of that right-hand 
portion of the curve on the diagram, describing a chiral first-order phase 
transition. The form of this curve reflects the fact that polarization 
screening effects become stronger with increasing $N_f$, and therefore 
the value of $g^{(0)}$, at which the first-order chiral phase transition 
occurs, increases with $N_f$. Note that it is called a bulk phase 
transition in the literature. 

At last, the left-hand portion of the curve, separating symmetric and 
non-symmetric phases, coincides with the line of the infrared-stable 
fixed points $\alpha^* (N_f)$ in Eq. (\ref{92}). It separates 
the symmetric, Coulomb, phase describing interactions of massless gluons 
and fermions, and the non-symmetric (and confinement) phase. 
Since it is a line of infrared-stable fixed points, it describes 
a first-order phase transition. 

Thus, in the Banks-Zaks picture, spontaneous chiral symmetry breaking 
and confinement occur in the weak-coupling phase at all $N_f < N_f^{**}
=\frac{11 N_c}{2}$. 
Notice that the left-hand part of the curve in the 
phase diagram describes a rather unusual situation: 
at $N_f^* < N_f < N_f^{**}$ spontaneous chiral symmetry breaking and 
confinement disappear with $\underline{\mbox{increasing}}$ the bare coupling 
constant $g^{(0)}$. 

As we shall show, the phase diagram changes dramatically if one adopts 
the suggestion of Ref. \cite{5}
concerning the dynamics of chiral symmetry breaking  
in this model. The suggestion is that since the value of the 
infrared-stable fixed point $\alpha^*$ is small at $N_f \simeq
N_f^{**} = \frac{11 N_c}{2}$, one should expect that there is a critical 
value of $N_f$, $N_f = N_f^{cr}$, above which the chiral symmetry is restored. 

In order to estimate the critical value $N_f^{cr}$, the authors of 
Ref. \cite{5} use the dynamical picture of chiral symmetry breaking 
corresponding to the rainbow (ladder) approximation. 
As in known \cite{10}, in a $SU(N_c)$ gauge theory, this picture coincides 
with that in quenched QED4, with the replacement the coupling constant 
$\alpha $ by $\alpha_{eff} = \frac{N_c^2 -1}{2N_c}\alpha$. 

Therefore, in this approximation, spontaneous chiral symmetry occurs 
when the gauge coupling exceeds a critical value 
$\alpha_c = \frac{2N_c}{N_c^2 -1} \cdot \frac{\pi}{3}$. 
Then, the estimate for the critical value $N_f^{cr}$ is 
\begin{equation}
\alpha^* (N_f) \left. \right|_{N_f = N_f^{cr}} = \alpha_c  \label{93}
\end{equation}
where $\alpha^* (N_f)$ is $\alpha^* (N_f) = - \frac{b}{c}$ with 
$b$ and $c$ from Eq. (\ref{91}). 

The estimate (\ref{93}) leads to the critical value
\begin{equation}
N_f^{cr} = N_c \left( \frac{100 N_c^2 - 66}{25 N_c^2 -15} \right).  \label{94}
\end{equation}
For $N_c = 3$, for example, $N_f^{cr}$ is just below 12.
\footnote{
As to a justification of this approximation, a computation of the 
next-to-leading term in gap equation shows that it yields a 
correction to $\alpha_c$ of approximately 
$\epsilon = \frac{1}{6 (1-\frac{1}{N_c^2})}$ \cite{31}. 
For $N_c=3$, the factor $\epsilon$ is $\epsilon = 0.19$. 
Therefore, if this factor reflects the contribution of higher orders, 
the estimate (\ref{93}) may be reliable. 
}. 

When $N_f^{cr} \leq N_f < N_f^{**} = \frac{11 N_c}{2}$, the value of the 
infrared-stable fixed point $\alpha^*$ is less than the critical value 
$\alpha_c$ and there is no chiral symmetry breaking.
There are two possibility: 
the bare coupling constant $\alpha^{(0)}$ is $\alpha^{(0)} \leq 
\alpha^{*}$ or $\alpha^{(0)} > \alpha^* $.

Let us first consider the case with $\alpha^{(0)} \leq \alpha^*$. 
Then, if $\alpha^{(0)} = \alpha^*$, the running gauge coupling $\alpha(\mu)$ 
is equal to the value $\alpha^*$ for all $\mu \leq \Lambda$. 
Then, as $\Lambda \to \infty$, one gets a conformal theory describing 
interactions of massless fermions and gluons. 

On the other hand, if $\alpha^{(0)} < \alpha^*$, the running coupling 
$\alpha (\mu)$ changes from $\alpha (\mu ) = \alpha^{(0)}$ at 
$\mu = \Lambda$ through $\alpha (\mu) = \alpha^*$ at $\mu = 0$. 
Nonperturbative effects, such as chiral symmetry breaking, are 
now absent, though the conformal symmetry is broken by ordinary 
perturbative contributions leading to running of $\alpha (\mu )$. 
Thus, in this case, an interacting non-Abelian Coulomb phase 
of massless quarks and gluons is still realized. 

Let us now consider the case with $\alpha^{(0)} > \alpha^*$. 

If the value of $\alpha^{(0)}$ is close to $\alpha^*$, then
the interactions are 
still weak, and chiral symmetry is unbroken. 
Therefore there is still an interacting Coulomb phase in this case, 
though, unlike the case with $\alpha^{(0)} < \alpha^*$, the running coupling 
decreases with $\mu$, and $\alpha (\mu) \to \alpha^*$ as $\mu \to 0$. 
As $\Lambda \to \infty$, we recover the conformal theory, with 
$\alpha (\mu ) = \alpha^*$ for all $\mu$, discussed above. 

As $\alpha^{(0)}$ becomes sufficiently large, one comes again to 
the first-order chiral (bulk) phase transition. 
The above consideration leads us to suggesting the phase diagram 
shown in Fig. 3. 

Notice that, as before, the form of the right-hand part of the curve, 
describing the bulk phase transition, reflects the point that the polarization screening effects becomes stronger with increasing $N_f$. 
In particular, the rainbow approximation, used at $N_f \simeq N_f^{cr}$, 
ceases to be good at larger $N_f$. 

The left-hand portion of the curve in Fig. 3 still coincides with 
the line of the infrared-stable fixed points $\alpha^* (N_f)$ in Eq. 
(\ref{92}). However, now it separates two symmetric (with unbroken chiral 
symmetry) phases, and, besides that, its lower end point is 
$N_f = N_f^{cr}$ 
and not $N_f = N_f^{*}$ as in Fig. 2: at $N_f < N_{f}^{cr}$
the infrared-stable fixed point is washed out by generating a
dynamical fermion mass. These two symmetric phases are 
distinguished by their dynamics at short distances: 
while the dynamics of one phase is asymptotically free, another is not. 
On the other hand, their long distances dynamics, governed by 
the infrared-stable fixed point, are similar. 

At last, the horizontal, $N_f = N_f^{cr}$, line describes the CPT-like 
phase transition in this model. The relation (\ref{74}) suggests the 
following scaling law for $m_{dyn}^2$:
\begin{equation}
m_{dyn}^2 \sim \Lambda_{cr}^2 \exp \left( - \frac{C}{
(\frac{\alpha^* (N_f)}{\alpha_c} -1)^{1/2}} \right) \; ,   \label{95}
\end{equation}
where $C$ is some constant $(C = 2\pi$ in Eq. (\ref{74})) and 
$\Lambda_{cr}$ is a scale at which the running coupling is of
order $\alpha_c$. 

The dynamics in a $SU(N_c) $ gauge theory with $N_f \simeq 
N_f^{cr}$ may be relevant for the realization of the scenario of 
``walking''-like technicolor for electroweak symmetry breaking \cite{32}: 
the ``walking''coupling constant $\alpha (\mu) \simeq \alpha^* \simeq 
\alpha_c$ governs the chiral symmetry breaking dynamics. 
In this case the effective action in quenched QED4, considered in Sec.4,
should be relevant for 
the description of this dynamics. 

Let us now turn to data of lattice computer simulations of a $SU(N_c)$ 
gauge theory. 

Lattice computer simulations of the $SU(3)$ theory with $N_f =8 $ and 
$N_f = 12$ of staggered fermions in Ref. \cite{33} show the presence 
of the first-order, bulk transition separating strong- and weak-coupling 
phases. While at $N_f = 12$ there is a clear signature of the chiral phase 
transition at which the order parameter jumps to zero, at $N_f =8$ 
the situation is less clear. 
More recent, and refined, simulations of this theory with $N_f =8$ 
indicates that in this case there is a first-order phase transition 
at which the order parameter jumps without vanishing \cite{34}. 

Thus the data of both these simulations seem indicate on
the presence of the 
vertical line (at least at $N_f \geq 8$) shown in Fig. 2 and Fig. 3. 
It is still impossible at present to discriminate between these two phase 
diagrams. 

Note that since the bulk phase transition is a lattice artifact, the form 
of the phase diagram can depend on the type of
fermions used in the simulations. 
The simulations of the $SU(3)$ theory with Wilson fermions \cite{35}
show that 
theories which satisfy both quark confinement and spontaneous 
chiral symmetry breaking in the continuum limit exist 
only for $N_f \leq 6$. 
When $16 \geq N_f \geq 7$, the theory is non-trivial (interacting), 
however, without quark confinement. Moreover, 
at $N_f \geq 7$, chiral symmetry is unbroken at all values of 
the bare coupling $\alpha^{(0)}$. 

These data seem favor the phase diagram in Fig. 3 in which the 
right-hand part of the curve is replaced by the horizontal line 
$N_f = N_f^{cr} = 7$. 

It is clear that more data are needed in order to establish firmly 
the phase diagram in a $SU(N_c)$ gauge theory.

\section{Conclusion}

In this paper we introduced the conception of the conformal phase 
transition (CPT) which provides a useful framework for studying 
nonperturbative dynamics in gauge (and also other) field theories. 
We described the general features of this phase transition. 

The CPT is intimately connected with the nonperturbative breakdown 
of the conformal symmetry, in particular, with the PCDC dynamics. 
In the non-symmetric phase the conformal symmetry is broken by 
marginal operators. This in turn yields a constraint on the form 
of the effective action in theories with the CPT.

In all the examples of the CPT considered in this paper, the 
conformal symmetry was explicitly broken by the conformal anomaly
in the phase with spontaneous chiral symmetry breaking. Is it
possible to realize dynamics with both chiral and conformal
symmetries being broken spontaneously? Although at present this
question is still open, we would like to note that long ago 
arguments had been given against the realization of such a
possibility \cite{42} 

The conception of the CPT can be useful for strong-coupling gauge theories, 
in particular, for QCD and models of dynamical electroweak symmetry 
breaking. 
In connection with that, we note that the effective action considered 
in Sec. 6 may be relevant for the description of $\sigma$ meson 
$( f^0 (400 - 1200))$ \cite{36,37}. If it is rather light 
(with $M_{\sigma} \simeq 600$ MeV) as some authors conclude \cite{37}, 
it can dominate in the matrix elements of the operator 
$\theta_{\mu}^{\mu}$ in low energy dynamics, i. e., it can be considered 
as a massive dilaton, as was already suggested some time 
ago \cite{25,27}
\footnote{
For a recent application of this conception in nuclear physics, see 
Ref. \cite{38}.
}.

It is also clear that the conception of PCDC and massive dilaton can 
be useful for the description of the dynamics of composite Higgs boson. 

Another application of the CPT (or pseudo-CPT)
may be connected with non-perturbative dynamics 
in condensed matter. Here we only mention the dynamics of non-fermi 
liquid which might be relevant for high-temperature surperconductivity: 
some authors have suggested that QED3 may serve as an effective theory of 
such a dynamics \cite{39}.

There has been recently a breakthrough in understanding non-perturbative
infrared dynamics in supersymmetric (SUSY) theories
(for a review see Ref. \cite{40} ). It would
be worth considering the realization of the CPT, if any, in SUSY theories,
thus possibly establishing a connection between SUSY and non-SUSY dynamics.

\section*{Acknowledgments}

V.A.M. thanks V. P. Gusynin and I. A. Shovkovy for useful discussions. He
wishes to acknowledge the JSPS (Japan Society for the Promotion of Science)
for its support during his stay at Nagoya University. This work is
supported in part by a Grant-in Aid for Scientific Research from the
Ministry of Education, Science and Culture (No. 08640365).

\section*{Appendix}
\renewcommand{\theequation}{\mbox{A.}\arabic{equation}}
\newcommand{\my}{\setcounter{equation}{0}}
\my

In this Appendix the relation for the conformal anomaly is derived. 

Since the relation for the conformal anomaly is an operator one, 
one can consider its realization for any matrix element. 

We will consider the vacuum expectation value 
$\langle 0 | \theta^{\mu}_{\mu} | 0 \rangle$
\footnote{
We define $\theta_{\mu}^{\mu}$ as 
$(\theta_{\mu}^{\mu})_{can} - \langle 0 | 
(\theta^{\mu}_{\mu})_{can} | 0 \rangle_0$, 
where $(\theta_{\mu}^{\mu})_{can} $ is the canonical 
expression for the trace of the energy-momentum tensor and 
$\langle 0 | (\theta^{\mu}_{\mu})_{can} | 0 \rangle_0 $ is its 
vacuum expectation value in the perturbative (massless) vacuum. 
For the models in question, this definition guarantees the finiteness 
of $\langle 0 | \theta^{\mu}_{\mu} | 0 \rangle$ in the 
continuum limit, after renormalizations of coupling constants. 
}. 
Then, because of the Lorentz invariance, 
$\langle 0 | \theta^{\mu}_{\mu} | 0 \rangle = 
4 \langle 0 | \theta^{0}_{0} | 0 \rangle = 4 \epsilon_V $, where 
$\epsilon_V$ is the vacuum energy density. 
It has the form: 
\begin{equation}
\epsilon_V = \Lambda^4 f( \{ g_i^{(0)} \} ),  \label{a-1}
\end{equation}
where $f$ is some function of dimensionless coupling constants $g_i^{(0)}$. 
Let us assume that the renormalization of $g_i^{(0)}$ leads to 
a finite $\epsilon_V$ in the continuum limit. 
This implies that 
\begin{equation}
\frac{d \epsilon_V}{d \ln \Lambda} = 4 \epsilon_V + 
\Lambda^4 \sum_i \frac{\partial f}{\partial g_i^{(0)}} 
\beta_i ( \{ g_i^{(0)} \} ) = 0 , \label{a-2}
\end{equation}
where $\beta_i = \frac{\partial g_i^{(0)}}{\partial \ln \Lambda}$. 
Therefore 
\begin{equation}
\langle 0 | \theta^{\mu}_{\mu} | 0 \rangle = 4 \epsilon_V = 
- \Lambda^4 \sum_i \frac{\partial f}{\partial g_i^{(0)}} 
\beta_i ( \{ g_i^{(0)} \} ) = 
- \sum_i  \frac{\partial \epsilon_V}{\partial g_i^{(0)}} \beta_i
( \{ g_i^{(0)} \} ).  \label{a-3}
\end{equation}
Then, using the path integral representation for $\epsilon_V$, we 
obtain the relations for the conformal anomaly considered in Secs. 3
and 4.

\newpage

\section*{Figure Captions}

Figure 1. The phase diagram in the gauged NJL model. The marks S and A
denote symmetric 
and asymmetric phases, respectively. 

\vspace{1cm}

Figure 2. The phase diagram in a $SU(N_c)$ gauge model suggested by 
Banks and Zaks. The marks S and A denote symmetric 
and asymmetric phases, respectively. 

\vspace{1cm}

Figure 3. The modified phase diagram in a $SU(N_c)$ gauge model discussed 
in the text. The marks S and A denote symmetric 
and asymmetric phases, respectively.

\newpage

\epsfbox{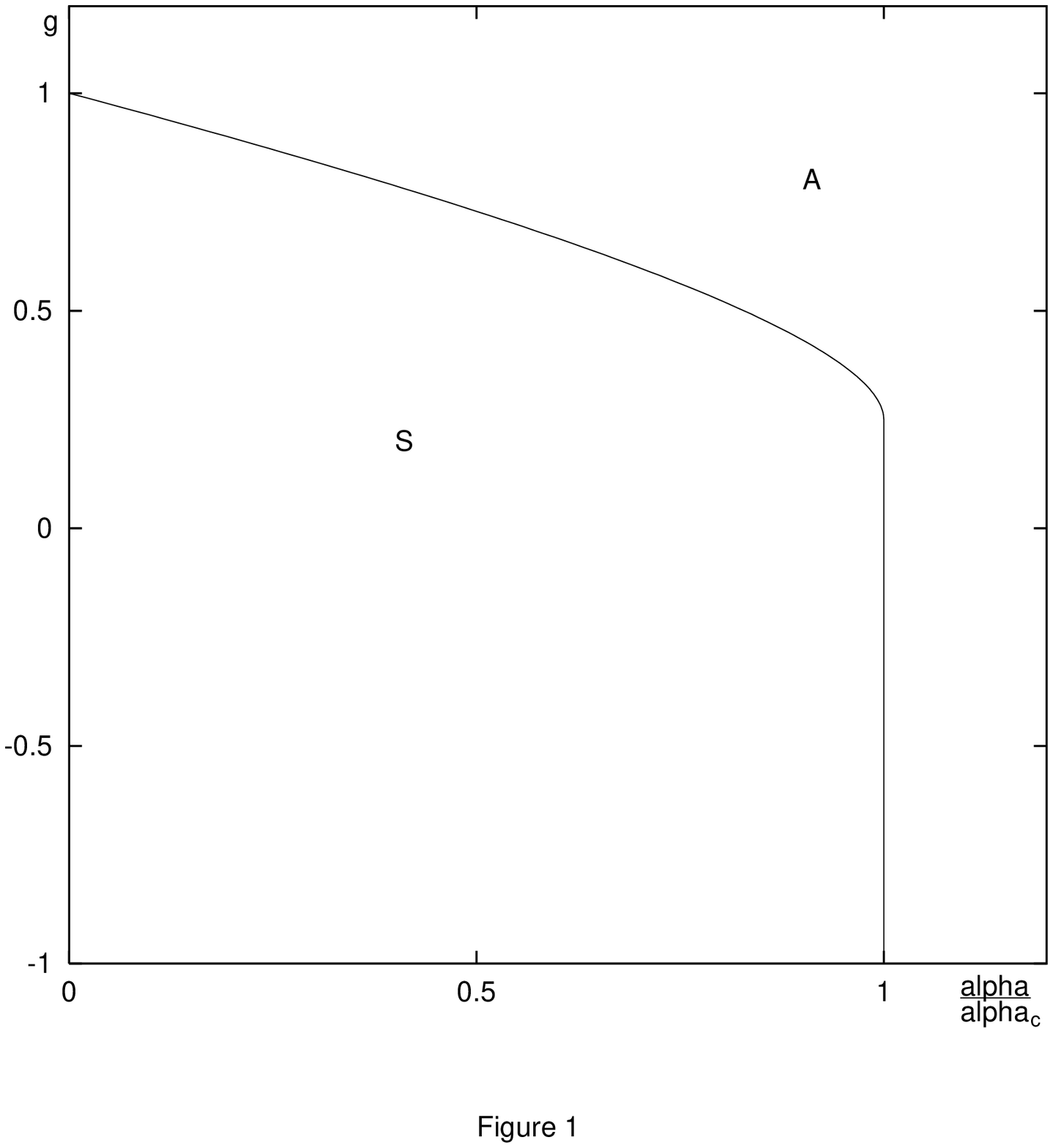}

\newpage

\epsfbox{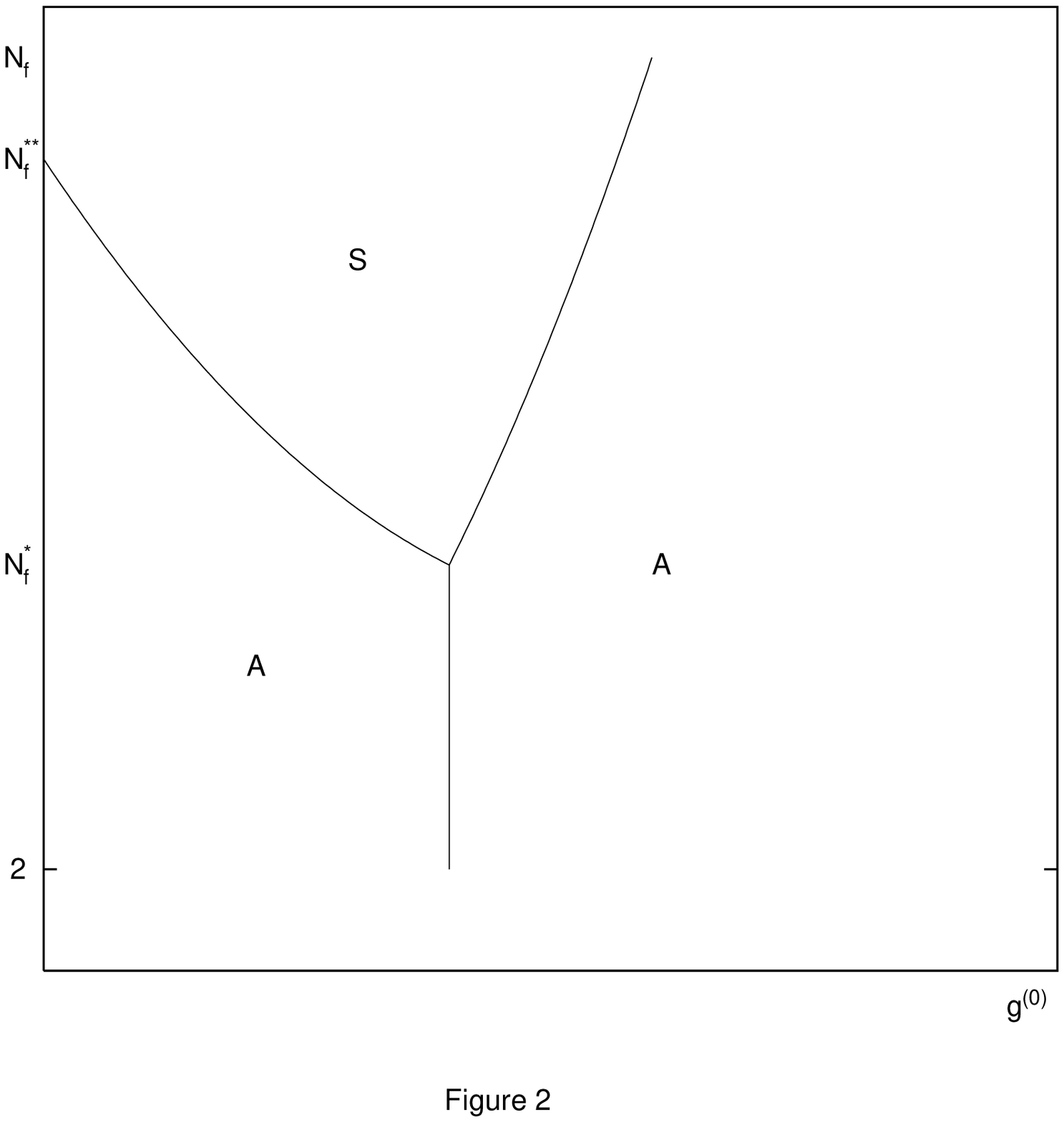}

\newpage

\epsfbox{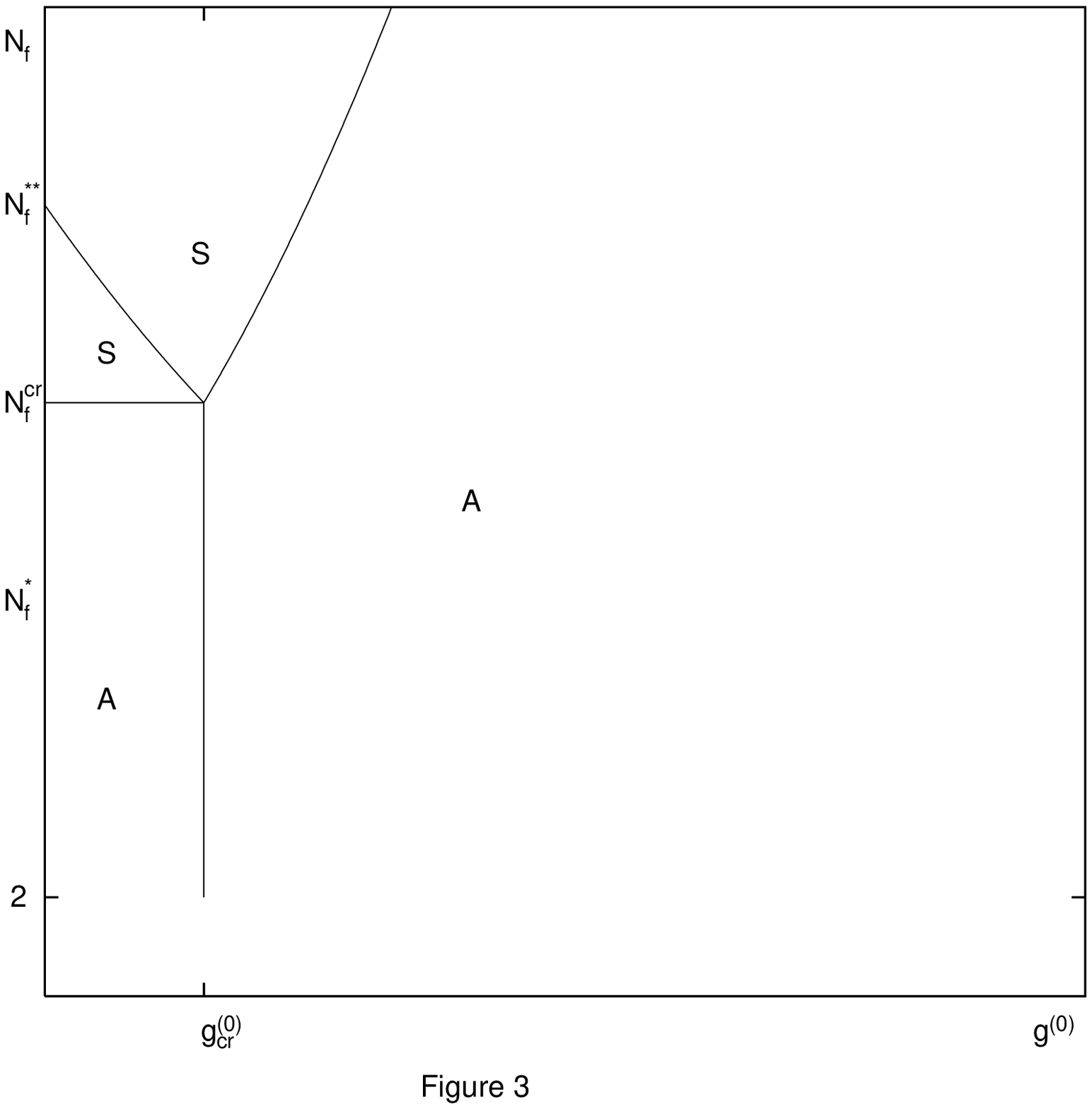}

\end{document}